\documentclass[12pt,a4paper]{article}

\usepackage{a4wide}
\usepackage{fancyhdr}
\usepackage{graphicx}
\usepackage{adjustbox}
\usepackage{caption}
\usepackage{subcaption}
\usepackage[ngerman, english]{babel}
\usepackage[utf8]{inputenc}
\usepackage[T1]{fontenc}
\usepackage{lmodern}
\usepackage{xcolor}
\usepackage[colorlinks,
citecolor={blue!70!black},
urlcolor={blue!70!black},
linkcolor={red!70!black},
hyperindex,breaklinks]{hyperref}
\usepackage{cite}
\usepackage{amsmath, amsthm, amssymb}
\usepackage{braket}
\usepackage{authblk}
\usepackage{tikz}
\usepackage{float}
\usepackage{indentfirst}
\usepackage[compat=1.1.0]{tikz-feynman}
\usepackage{diagbox}

\title{How Special Are Black Holes? \\
{\Large Correspondence with saturons in generic theories}}

\author{Gia Dvali\thanks{georgi.dvali@physik.uni-muenchen.de} $^{1,2}$, Oleg Kaikov\thanks{Kaikov.Oleg@physik.uni-muenchen.de} $^{1,2}$, and Juan Sebasti\'{a}n Valbuena Berm\'{u}dez\thanks{juan.valbuena@physik.lmu.de} $^{1,2}$\\
$^1${\small{\em Arnold Sommerfeld Center, Ludwig-Maximilians-Universit{\"a}t,
}}\\
{\small{\em Theresienstra{\ss}e 37, 80333 M{\"u}nchen, Germany
}}\\
$^2${\small{\em
Max-Planck-Institut f{\"u}r Physik,}}\\
{\small{\em F{\"o}hringer Ring 6, 80805 M{\"u}nchen, Germany
}}\\
}

\date{\small{\today}}

\begin{document}

\maketitle

\begin{abstract}
\noindent
Black holes are considered to be exceptional 
due to their time evolution and information processing. However, it was proposed recently that these properties 
are generic for objects, the so-called \emph{saturons}, that attain the maximal entropy permitted by unitarity. In the present paper, we verify this connection within a renormalizable $SU(N)$ invariant theory.
We show that the spectrum of the theory contains a tower of bubbles representing bound states of $SU(N)$ Goldstones. Despite the absence of gravity,  
a saturated bound state  
exhibits a striking correspondence with a black hole: Its entropy is given by the Bekenstein-Hawking formula; semi-classically, the bubble evaporates at a thermal rate with a temperature equal to its inverse radius;  the information retrieval time is equal to Page's time. 
The correspondence goes through a trans-theoretic entity of Poincar\'{e} Goldstone. 
The black hole/saturon correspondence has important implications for black hole physics, both fundamental and observational.

\end{abstract}

\section{Introduction}

Black holes are considered to be exceptional objects. This exceptionality usually refers to their time-evolution, information storage, and retrieval. However, it has been proposed recently \cite{Dvali:2020wqi} (see \cite{Dvali:2021jto} for a summary) that these properties are not specific to black holes. Rather, they are the universal features of objects saturating a certain bound on the micro-state entropy that shall be discussed below. This bound is imposed by unitarity. 

This proposal heavily relied on the evidence gathered in two earlier papers \cite{Dvali:2019ulr, Dvali:2019jjw}. There, objects such as solitons, instantons, and various bounds states  with high degeneracy of micro-states, were constructed.  
This gave a possibility of increasing the micro-state entropy in a controllable way while simultaneously monitoring its correlation with unitarity.  
In all cases, this correlation pointed to a universal bound on entropy.\\
  
The bound can be presented in two equivalent forms \cite{Dvali:2020wqi}. The first form states that for an arbitrary self-sustained object of size $R$ in $d$ space-time dimensions, the unitarity imposes the following bound on entropy,  
\begin{equation} \label{Area}
S \leqslant \dfrac{\mathcal Area}{G_{\text{Gold}}}\, ,     
\end{equation}  
where ${\mathcal Area} \sim R^{d-2}$ is the area of the sphere within which the object is contained, and $G_{\text{Gold}}$ is the coupling of the Goldstone field of a spontaneously broken symmetry. Such spontaneously broken symmetries are always present whenever we are dealing with a macroscopic object of high entropy. In particular, any ``device'' capable of storing quantum information breaks the Poincar\'{e} symmetry spontaneously. Therefore, no ambiguity exists in defining $G_{\text{Gold}}$. Note, the bound is valid in arbitrary dimensions, including $d=2$, since the quantity ``${\mathcal Area}$'' is well defined. \\
   
The second form of the entropy bound (\ref{Area}) is written in terms of an effective coupling $\alpha$ of the theory. In terms of this coupling, the bound imposed by unitarity can be written in the following form, 
\begin{equation} \label{alphaB}
S \leqslant \dfrac{1}{\alpha} \, ,
\end{equation}
where $\alpha$ has to be understood as a running coupling 
evaluated at the scale $1/R$. The above bound is equivalent to (\ref{Area}). This can already be noticed from the fact that the quantity
\begin{equation}
\alpha_{\text{Gold}} \equiv \dfrac{G_{\text{Gold}}}{\mathcal Area}
\end{equation}
represents an effective dimensionless coupling of the Goldstone
boson, evaluated at the scale $1/R$. Both bounds are saturated simultaneously \cite{Dvali:2019ulr}. Correspondingly the maximal entropy compatible with unitarity can be written in the combined form, 
\begin{equation} \label{Smax}
S_{\text{max}} = \dfrac{1}{\alpha}= \dfrac{\mathcal Area}{G_{\text{Gold}}} \, ,
\end{equation}
The coupling $G_{\text{Gold}}$ can be expressed in terms of the canonically normalized Goldstone decay constant $f$ as, 
\begin{equation}
G_{\text{Gold}} \equiv f^{-2} \, .
\end{equation}
In this normalization, in $d=4$, in which we shall work in the present paper, $f$ has the dimensionality of mass. For any self-sustained bound-states of $N$ quanta of wavelengths $R$, the coupling and the decay constant of a Poincar\'{e} Goldstone are given by,
\begin{equation} \label{Gdecayf}
G_{\text{Gold}} = f^{-2} = \dfrac{R^2}{N} \, .
\end{equation}

The objects saturating the entropy bounds (\ref{Area}) and (\ref{alphaB}) shall be referred to as ``saturons''. Now, the point of \cite{Dvali:2020wqi} is that saturons share the following universal properties: 
\begin{itemize}
  \item Of course, as already said, their entropy satisfies the area law, given by (\ref{Smax}).
  \item If unstable, they decay with the rate which, up to 
  $1/S$ corrections, gives the thermal rate of temperature 
  \begin{equation} \label{Temp}
  T \sim 1/R \, .
 \end{equation} 
 \item In semi-classical treatment, they exhibit an information horizon.
   \item The minimal time-scale required for the start of the information retrieval is bounded from below by,
    \begin{equation} \label{Volume}
t_{\text{min}} = \dfrac{{\mathcal Volume}}{G_{\text{Gold}}} = 
\dfrac{R}{\alpha} = S_{\text{max}} R \, ,
\end{equation}
where ${\mathcal Volume} \sim R^{d-1}$.
\end{itemize} 
  
It is obvious that the above properties are in full correspondence with the properties of black holes, under the mapping,
\begin{equation} \label{GandG}
     G_{\text{Gold}} \, \rightarrow \,  G_{\text{N}} \, , 
\end{equation}
where $G_{\text{N}}$ is Newton's gravitational constant. The fundamental meaning of this mapping is that in the case of a black hole, the Goldstone boson of spontaneously broken Poincar\'{e} symmetry comes from the graviton, which has the coupling  $G_{\text{N}}$. \\
   
It is clear that under the transformation (\ref{GandG}) the entropy (\ref{Smax}) becomes the Bekenstein-Hawking entropy \cite{Bekenstein:1973ur, Hawking:1975vcx} of a black hole of radius $R$,
 \begin{equation}
   S_{\text{BH}} \sim \dfrac{\mathcal Area}{G_{\text{N}}} \, .  
\end{equation}
We also notice that the same entropy can be written as \cite{ Dvali:2020wqi,Dvali:2011aa},
  \begin{equation}
	S_{\text{BH}} \sim \dfrac{1}{\alpha_{\text{gr}}} \, ,
\end{equation}
where  $\alpha_{\text{gr}} = G_{\text{N}}/{\mathcal Area}$ is the quantum gravitational coupling in $d$ space-time dimensions, evaluated at the scale of the momentum transfer $1/R$.
    
Next, the connection  between (\ref{Temp}) and Hawking radiation is very transparent. Finally, we notice that under the mapping (\ref{GandG}), the time-scale (\ref{Volume}) gives a so-called Page's time for a black hole \cite{Page:1993wv}. For a black hole, this time was suggested to be the minimal time-scale of the start of information recovery. \\
   
The proposal of \cite{ Dvali:2020wqi} says that none of the above properties are specific either to black holes or to gravity. Rather, they are generic features of saturons in arbitrary theories. 
 Various aspects of this connection have been demonstrated for  
 solutions in spaces both with Lorentzian and Euclidean signatures, such as solitons, instantons, and other bound states \cite{Dvali:2019ulr, Dvali:2019jjw, Dvali:2020wqi}, 
including \cite{Dvali:2021rlf} bound states in the Gross-Neveu model \cite{Gross:1974jv} on which we shall comment  below. 
 Recently, it has also been argued \cite{Dvali:2021ooc}
 that a so-called ``color glass condensate'' state of ordinary QCD \cite{Gelis:2010nm} exhibits the same 
 correspondence. \\

The correspondence between black holes and other saturons should not be understood as the correspondence between different theories. Rather, the correspondence is between the states in different theories. The starting point which these states share is that they all saturate the entropy bounds (\ref{Area}) and (\ref{alphaB}), imposed by unitarity of corresponding theories. The rest of the features follow. In other words, saturation is a trans-theoretic notion. \\
       
The correspondence between black holes and generic saturons offers an obvious benefit of ``demystifying'' black hole physics. Indeed, on the one hand, saturons exist in theories that are renormalizable and calculable. On the other hand, they exhibit all of the key properties of a black hole. Hence, by studying saturons, we may explain the origin of known black hole properties and, at the same time, predict new phenomena that can take place in black holes. One such effect is the phenomenon of ``memory burden'' \cite{Dvali:2018xpy, Dvali_2020_2}.  The essence of it is that quantum information can stabilize the system that carries this information. \\     
      
In the present paper, we shall provide further evidence of the black hole/saturon correspondence. We shall analyze a simple four-dimensional theory of saturons, constructed in \cite{Dvali:2020wqi}. The model exhibits a global $SU(N)$ symmetry and has several degenerate vacuum states with various patterns of spontaneous symmetry breaking. The theory is renormalizable and can be studied at arbitrarily weak coupling and large $N$. The spectrum of the theory can be reliably analyzed. It contains solitonic vacuum bubbles separating the vacua with different patterns of broken symmetry.  
  
In particular, there exists an infinite set of bubbles stabilized by the memory burden effect. Each bubble represents a bound state of a large number of Goldstone bosons. The occupation number of the Goldstone quanta in the bound state will be denoted by $N_G$. These quanta cannot propagate in the asymptotic vacuum outside a bubble since the symmetry is unbroken there. Due to this, the Goldstones can only exist within a bubble. Therefore,  if a bubble decays, the conserved $SU(N)$ quantum numbers carried by the Goldstone modes must be released in the form of massive quanta. This creates an energy barrier that stabilizes the bubble in the semi-classical theory. 

The levels of the spectrum can be labeled by $N_G$. Each level is exponentially degenerate. Correspondingly, the bound states carry a high micro-state entropy. We shall see that the bubbles with occupation number $N_G \sim N$ are saturons. In particular, the entropy of such a bubble saturates the bounds (\ref{Area}) and (\ref{alphaB}). Correspondingly, they have maximal entropy (\ref{Smax}) compatible with unitarity. We shall show that such a bound state assumes all properties of a black hole listed above. We thus confirm the black hole / saturon correspondence proposed in \cite{Dvali:2020wqi}. \\
   
The fact that a saturated bound state of $N$ Goldstone bosons exhibits a correspondence with a black hole represents supporting evidence for the idea of the black hole $N$-portrait \cite{Dvali:2011aa}. In this theory, the black hole is described as a saturated bound state of $N$ gravitons.  The features predicted by this theory of gravitons are explicitly reproduced in a calculable renormalizable theory of the Goldstone bound states presented in the present paper.  In particular, we observe how the information is carried away by $1/S$ ($1/N$) corrections to Hawking 
radiation. The same corrections violate the  self-similarity of the evaporation process.
\\
   
Finally, we wish to say that in a parallel paper \cite{Dvali:2021rlf}, the black hole saturon correspondence was demonstrated within the Gross-Neveu model \cite{Gross:1974jv, Dashen:1975xh} (for a review, see \cite{Coleman:1980nk, Manohar:1998xv}). 
 This theory contains a set of fermions-bound states with growing mass and degeneracy. It was shown \cite{Dvali:2021rlf} that the Gross-Neveu bound state of the maximal degeneracy is a saturon and exhibits all properties of a black hole discussed above. This is remarkable. Despite the fact that the Gross-Neveu model is two-dimensional, its saturon captures all properties of saturons (and correspondingly of black holes) of higher dimensional theories. This includes the saturons in the four-dimensional $SU(N)$ symmetric model discussed in the present paper. This fact illustrates the universality of large-$N$ physics at the point of saturation.  This is the reason why saturated fermion bound states in Gross-Neveu and saturated Goldstone bound states in the present model share strong similarities with each other and with black holes. This fact provides non-trivial supporting evidence to the idea that black hole properties are not specific to gravity but instead are defined by the generic  
physics of saturation \cite{Dvali:2020wqi}.

\section{A model of a saturon}

\subsection{The model} 

Following \cite{Dvali:2020wqi}, we consider a theory of a scalar field $\phi$ in the adjoint representation of $SU(N)$, with $N \geq 3$. That is, $\phi_{\alpha}^{\beta}$ is an $N \times N$ traceless Hermitian matrix with $\alpha, \beta = 1, \dotsc, N$. The Lagrangian density is

\begin{eqnarray}
\label{model_vac_bub}
\mathcal{L} & = & \dfrac{1}{2} \text{tr} \left[ \left( \partial_{\mu} \phi \right) \left( \partial^{\mu} \phi \right) \right] - V\left[ \phi \right] \,, \\ \nonumber
{\rm where}, \quad V\left[ \phi \right] &=& \dfrac{\alpha}{2} \text{tr} \left[ \left( f \phi - \phi^2 + \dfrac{I}{N} \text{tr} \left[ \phi^2 \right] \right)^2 \right] \, .
\end{eqnarray}

Here, $I$ is the unit $N \times N$ matrix, $\alpha$ is a dimensionless coupling and $f$ is the scale of symmetry breaking.  

Notice, the theory is renormalizable. We shall restrict ourselves to the regime in which the fundamental quantum coupling $\alpha$ is weak. However, even for arbitrarily-weak $\alpha$, the unitarity imposes the following upper bound on the strength of the coupling, 
\begin{equation} \label{Ubound}
\alpha \lesssim \dfrac{1}{N} \, .
\end{equation} 
This bound is fully non-perturbative. The physical reason is that the parameter that controls unitarity, both perturbatively as well as  non-perturbatively, is the collective coupling $\alpha N$. In the context of gauge theories, such coupling is often referred to as the 't Hooft coupling \cite{tHooft:1973alw}. \\

Regardless of smallness of $\alpha$, the unitarity is saturated when the collective coupling $\alpha N$
becomes order one. This is signalled both 
by the breakdown of loop-expansion as well as by 
saturation of unitarity by the scattering amplitudes. 
For the detailed discussion we refer the reader to
\cite{Dvali:2020wqi}. 
As shown in this work, the constraint (\ref{Ubound}) plays a fundamental role in enforcing the upper bound on entropy (\ref{Smax}). We shall present an explicit example of saturation of this bound in the $SU(N)$ theory (\ref{model_vac_bub}). \\
   
We shall work in the regime of weak $\alpha$ and large $N$. Interesting things happen when the collective coupling $\alpha N$ approaches the unitarity bound (\ref{Ubound}) from below. As we shall see, in this limit certain solitons saturate the entropy bounds (\ref{Area}) and (\ref{alphaB}) and start behaving like black holes. \\ 
   
In order to identify such solitons, let us notice that the vacuum equations,
\begin{equation}
f\phi^\beta_\alpha -(\phi^2)^\beta_\alpha +\dfrac{\delta^\beta_\alpha}{N} \text{tr} \left[\phi^2 \right] = 0,
\end{equation}
admit several solutions corresponding to vacua with different unbroken symmetries. These include the vacuum with unbroken $SU(N)$ symmetry, $\phi=0$, and the vacua with spontaneous symmetry breaking (SSB) patterns $SU(N) \rightarrow SU(N-K) \times SU(K) \times U(1)$, with $0<K<N$. By construction, all of these vacua are degenerate in energy. For definiteness, we  focus on the vacuum $\phi = 0$ and the one with $K = 1$.  In the second vacuum, only the following component
\begin{equation} \label{nonzerocomponent}
\phi_{\alpha}^{\beta} = \dfrac{\phi(x)}{\sqrt{N(N-1)}} \text{diag} \left( (N-1), -1, \dotsc, -1 \right) \, ,
\end{equation}
has a non-zero expectation value
\begin{equation} \label{VEV}
\braket{\phi}=f \dfrac{\sqrt{N(N-1)}}{(N-2)} \simeq f \, ,
\end{equation} 
(since we shall be working at large $N$, when appropriate, we shall approximate parameters by their leading order values). In this vacuum, the symmetry group is 
  $SU(N-1) \times U(1)_Y$, where the generator of 
  $U(1)_Y$ is, 
 \begin{equation}
\hat{Y} = \dfrac{1}{\sqrt{2N(N-1)}} \text{diag} \left( (N-1), -1, \dotsc, -1 \right) \, .
\end{equation}

In the vacuum with unbroken $SU(N)$ symmetry, there exist no massless excitations, as the theory in this vacuum exhibits a mass gap,
\begin{equation} \label{mass}
m = \sqrt{\alpha} f \, .
\end{equation}
In contrast, in the broken symmetry vacuum there exist 
\begin{equation}
N_{\text{Gold}} = 2(N-1)
\end{equation}
different species (``flavors'') of massless Goldstone bosons, which we denote by $\theta^a(x_{\mu})$ with $a=1,2, \dotsc , N_{\text{Gold}}$. They correspond to broken generators  $T^a$. It is convenient to represent these broken generators as off-diagonal Pauli matrices.
We can take
 \begin{equation}
(T^a)_{\alpha}^{\beta} = \dfrac{1}{2}(\delta_{\alpha}^1 \delta_{k+1}^{\beta} + \delta_{\alpha}^{k+1} \delta_1^{\beta}) \, \quad \text{for} ~ a = 2k -1 \, , 
\end{equation}
and 
\begin{equation}
(T^a)_{\alpha}^{\beta} = \dfrac{-i}{2}(\delta_{\alpha}^1\delta_{k+1}^{\beta} - \delta_{\alpha}^{k+1}\delta_1^{\beta}) \, \quad \text{for} ~ a= 2k \, ,
\end{equation} 
where $k =1,\dotsc,N-1$. 
 
The combinations, $T_k^{\pm} = T^{2k-1} \pm iT^{2k}$ form $(N-1)$-dimensional fundamental and anti-fundamental  representations under the unbroken $SU(N-1)$ symmetry group. Their charges $Y_{\pm} = \pm \frac{N}{\sqrt{2N(N-1)}}$ under the $U(1)_Y$-symmetry are fixed by  
\begin{equation}
[\hat{Y},T_k^{\pm}] =  Y_{\pm} T_k^{\pm} \, .
\end{equation}
Correspondingly, the combinations of Goldstones, $\theta_k^{\pm} \equiv  \theta^{2k-1} \pm i\theta^{2k}$, transform under the same representations as the generators $T_k^{\pm}$. Obviously, $\sum\limits_a \theta_a T^a = \frac{1}{2} \sum\limits_k \theta_k^{\mp} T_k^{\pm}$. \\
   
The effective low energy theory of the Goldstone modes can be obtained in the standard way, by parametrizing the field as 
\begin{equation}
\phi_{\alpha}^{\beta} = \left( U^{\dagger} \braket{\phi} U \right)_{\alpha}^{\beta} 
\end{equation}
with
\begin{equation}
\quad U = \exp \left[ - i \theta^a T^a \right] \,,
\end{equation}
and subtituting into the original action. Taking into account the commutation relations 
and keeping terms up to second order in $\theta^a$, we obtain the following effective Goldstone Lagrangian
 \begin{equation}
\mathcal{L}_{\text{eff}} =  \dfrac{N^2}{4(N-2)^2} f^2 \sum_a\left(\partial_{\mu} \theta^a \right) \left(\partial^{\mu} \theta^a \right) \simeq \dfrac{1}{4} f^2 \sum_k\left(\partial_{\mu} \theta_k^{+} \right) \left(\partial^{\mu} \theta_k^{-} \right) \, .
\end{equation}

Now, due to gapless Goldstone excitations, an ``island'' of a broken symmetry vacuum, embedded into a symmetric one, can carry a very high micro-state entropy. Such islands are realized in the form of the vacuum bubbles which we shall now consider.

\subsection{Vacuum bubble}  
 
Since the vacua  $SU(N)$ and $SU(N-1)\times U(1)_Y$ are degenerate, there exist domain walls that separate the two. A planar infinite wall is static.  For such a wall, the component (\ref{nonzerocomponent})
of the adjoint field has the following form,
\begin{equation}
\phi(x)= \dfrac{f}{2} \left[ 1 \pm \tanh \left( \dfrac{ mx}{2} \right) \right] \, ,
\end{equation}
where $x$ is a coordinate that is perpendicular to the wall. The thickness of the wall is
\begin{equation}
\delta_{\text{w}} \sim m^{-1} \, . 
\end{equation} 
Likewise, we can consider a closed bubble of radius $R \gg \delta_{\text{w}}$, inside of which the $SU(N)$ symmetry is broken down to $SU(N-1) \times U(1)_Y$.  Outside of the bubble, the $SU(N)$-symmetry is unbroken. The exact analytic solution for a finite radius bubble is not known. Moreover, such a bubble experiences the force of tension directed inwards. If not counteracted by an equal force, the finite size bubble will collapse. 
   
For a slow-moving bubble wall, in the regime $R \gg \delta_{\text{w}}$, the profile of the $\phi$ field can be approximated by 
\begin{equation}
\phi(r)= \dfrac{f}{2} \left[ 1 + \tanh \left( \dfrac{ m(R -r)}{2} \right) \right] \, .
\end{equation}
Adopting the standard terminology, we shall refer to this approximation as the ``thin-wall'' approximation. \\

Due to SSB of the $SU(N)$ symmetry to $SU(N-1) \times U(1)$, the vacuum in the interior of the bubble contains $N_{\text{Gold}} = 2(N-1)$ gapless Goldstone bosons. These Goldstone modes are ``trapped'' inside the bubble and cannot propagate outside. This is because in an asymptotic vacuum  outside of the bubble, $r \gg R$,  the symmetry is unbroken. There, the theory has a mass gap (\ref{mass}) and no Goldstone excitations can penetrate this region. 
The existence of gapless species in the bubble interior, 
endows the bubble with high capacity of information storage
\cite{Dvali:2020wqi}. 
The information can be stored in the excitations of the Goldstone modes.  This results in a memory burden effect that counteracts the collapse.

\section{Bubble stabilization by memory burden} 

We now wish to study how the effect of memory burden influences the evolution of the bubble. The quantum information can be encoded in excitations of Goldstone modes at relatively low energy cost. 
This quantum information can be parameterized in form 
of sequences of excitation levels (occupation numbers 
$n^a$) of the Goldstone 
bosons with distinct  $SU(N)$ flavor quantum numbers. 
 We shall refer to
distinct arrangements as the ``memory patterns''. 
 
When the bubble collapses and decays, this information must be carried away by the outgoing quanta. However, the information pattern, released in the $SU(N)$-symmetric vacuum, is much more costly in energy than when it is stored in $S(N-1) \times U(1)$-symmetric vacuum inside the bubble. This is because no gapless excitations exist in the $SU(N)$-symmetric vacuum. In fact, as we shall see, the energy cost of the pattern outside the bubble can be larger than the entire energy of the system. This creates a conflict. \\

Correspondingly, the information encoded in Goldstone modes must back react and create a resistance against the decay. That is, the occupied Goldstone modes should prevent the bubble from collapsing and decaying. The memory burden effect is measured by the overall excitation of the Goldstone modes.  We may choose to excite a large diversity of modes to low levels each, or highly excite a single mode. In the latter case, we can approximate a corresponding Goldstone mode with a classical field.   Below, we shall encounter both regimes. \\
 
Let us again consider the model (\ref{model_vac_bub}) with the 
field given by
\begin{equation} \label{ROTATION}
\phi_{\alpha}^{\beta} = \left( U^{\dagger} \Phi U \right)_{\alpha}^{\beta} \, ,
\end{equation}
where 
\begin{equation}
\Phi_{\alpha}^{\beta} = \dfrac{\varphi(\vec{x},t)}{\sqrt{N(N-1)}} \text{diag} \left( (N-1), -1, \dotsc, -1 \right) \, ,
\end{equation}
and
\begin{equation}
\quad U = \exp \left[ - i \theta^a(\vec{x},t) T^a \right] \, .
\end{equation}

Substituting the above ansatz into (\ref{model_vac_bub}), we obtain
\begin{equation} \label{LEFF}
\mathcal{L} = \dfrac{1}{2} \left(\partial_{\mu} \varphi \right) \left(\partial^{\mu} \varphi \right) + \dfrac{N}{4(N-1)} \varphi^2 \left(\partial_{\mu} \theta^a \right) \left(\partial^{\mu} \theta^a \right) - \dfrac{\alpha}{2} \varphi^2 \left(\varphi - f \right)^2 \, ,
\end{equation}
where we  have absorbed the $1/N$-corrections into the redefinitions of parameters $f$ and $\alpha$.

We assume that the field $\varphi$ takes the initial configuration of a finite-size bubble of the broken symmetry vacuum embedded in the symmetric one. For definiteness, we shall assume spherical symmetry. That is, at some initial time, the field is described by a configuration $\varphi(0,r)$ which interpolates from a non-zero value at $r=0$ to zero at $r = \infty$. As already said, such a bubble can carry a large amount of information stored in excitations of the Goldstone fields $\theta^a$. These form a memory pattern. If none of the Goldstone modes are excited, the information pattern is empty and cannot affect the dynamics of the bubble. In such a situation, the bubble collapses and after some oscillations decays into the asymptotic quanta of the 
$SU(N)$-symmetric vacuum.   

We now wish to study how a non-empty information pattern changes the story. Depending on the occupation numbers of the individual Goldstone flavors, we can distinguish the following two regimes.  
   
When the individual occupation numbers are large, we can use Bogoliubov  approximation and replace the corresponding field operators by $c$-numbers. This regime can be studied classically, by treating the highly excited Goldstone modes $\theta^a$ as classical fields. In the opposite case, when the individual occupation numbers are small, such a replacement is not possible and we need to study the Goldsone modes in a quantum regime. 

Notice that the quantum regime by no means implies that the memory burden is weak. Although the individual occupation numbers are small, the diversity of excited Goldstone flavors can be very high for large $N$. As a result, even a classical 
bubble can be stabilized by the memory burden
that is fully quantum. 
As we shall see, this can happen if the bubble saturates the entropy bounds (\ref{Area}) and (\ref{alphaB}). We shall discuss the classical regime first.   
      
\subsection{Classical regime} 

We shall look for a localized stationary solution in which we give the Goldstone modes a time-dependence. To start with, we use the ansatz,
\begin{equation} \label{ansatz}
\varphi = \varphi(r)\, , ~ \theta^a = \delta^{a1} \omega t \, ,  
\end{equation}
where $\omega$ is the frequency of rotation in the internal space. Only one of the Goldstone modes is involved in this rotation. In other words, as previously mentioned, we occupy macroscopically a single Goldstone flavor. \\
   
Since the Goldstones with labels $a \neq 1$ are all set to zero, the effective Lagrangian (\ref{LEFF}) describes a full non-linear classical theory for the ansatz (\ref{ansatz}). 
Notice, this is legitimate because all other components 
enter bi-linearly (or in higher powers) in the Lagrangian. 
It is therefore  consistent to set them to zero without 
affecting the validity of the ansatz
(\ref{ansatz})~\footnote{~For example, the effective Lagrangian obtained 
by the ansatz  (\ref{ROTATION})
in which the matrix $U =\begin{pmatrix}
 \cos(\frac{\theta^1}{2}) \mathrm{e}^{-i\alpha}    & - \sin(\frac{\theta^1}{2}) \mathrm{e}^{i\beta}   \\
  \sin(\frac{\theta^1}{2}) \mathrm{e}^{-i\beta}     &  \cos(\frac{\theta^1}{2}) \mathrm{e}^{i\alpha}
\end{pmatrix}$ represents a generic $SU(2)$-transformation 
of the subgroup $SU(2)\times SU(N-2) \times U(1) \subset SU(N)$,  
 would give an effective Lagrangian
(with the same rescaling of $f$ and $\alpha$ as before),  
\begin{equation*}
\mathcal{L} = \dfrac{1}{2} \left(\partial_{\mu} \varphi \right) \left(\partial^{\mu} \varphi \right) + \dfrac{N}{4(N-1)} \varphi^2 \left(\partial_{\mu} \theta^1 \right) \left(\partial^{\mu} \theta^1 \right) +  \dfrac{N \sin^2(\theta^1)}{4(N-1)} \varphi^2 \left(\partial_{\mu} (\alpha + \beta) \right)  \left(\partial^{\mu} (\alpha + \beta) \right) - \dfrac{\alpha}{2} \varphi^2 \left(\varphi - f \right)^2 \, ,
\end{equation*}
for which $\alpha + \beta =0$ is a solution. }. 

Correspondingly, the only non-trivial equation is,
 \begin{equation} \label{radial}
\mathrm{d}_r^2\varphi + \dfrac{2}{r} \mathrm{d}_r\varphi + 
\varphi \left(  \omega^2 - \alpha (\varphi - f)( 2 \varphi  - f)\right )  = 0 \, .
\end{equation}
Here we have again absorbed the $N$-dependent pre-factors into the redefinitions of $f$, $\alpha$ and
\begin{equation}
\sqrt{\dfrac{N}{2(N-1)}} \omega \to \omega \, .
\end{equation}
A stationary bubble is described by a solution with the following boundary conditions,
 \begin{equation} \label{BCofansatz}
\varphi(0) \neq 0 \, , ~ \varphi(\infty) = 0 \, .
\end{equation}
Notice, the Goldstone fields of primary symmetry breaking, $SU(N) \rightarrow SU(N-1)\times U(1)_Y$, transform non-trivially under $SU(N-1)\times U(1)_Y$. Due to this, the ansatz (\ref{ansatz}) induces a secondary symmetry breaking down to $SU(N-2)\times U(1)_X$, where the generator of $U(1)_X$ is given by,
 \begin{equation}
  \hat{X} = \dfrac{1}{\sqrt{N(N-2)}} \text{diag} \left( (N-2)/2, (N-2)/2,-1, \dotsc, -1 \right) \, .
 \end{equation}
Due to this secondary symmetry breaking, there emerges an additional set of Goldstone modes. Their number is $2N-3$. Note, however, that there is a mixing between some internal Goldstones with the Goldstone boson of broken time-translation symmetry. This is an important phenomenon but does not influence the entropy count for large $N$. \\
   
Let us focus on the stationary classical configuration. If we mentally think of $r$ as a time coordinate, the above equation is equivalent to the one satisfied by a ``coordinate'' $\varphi$ of a particle moving in the external potential
\begin{equation}
V(\varphi) = \dfrac{1}{2}\varphi^2  \left(  \omega^2 - \alpha  (\varphi - f)^2\right ) \, .
\end{equation}
 For 
 \begin{equation} \label{MinC}
 \dfrac{\omega^2}{\alpha f^2} \, < \, 1 \, , ~ \, \text{or equivalently}, \, ~ \omega^2 \, < \, m^2 \, ,
 \end{equation}
the above potential has two maxima, $\varphi_0 = 0, \, ~ \varphi_{\text{max}} = \frac{f}{4} (3 + \sqrt{ 1 + \frac{8\omega^2}{\alpha f^2}} )$ separated by a minimum at $\varphi_{\text{min}} = \frac{f}{4} (3 - \sqrt{ 1 + \frac{8\omega^2}{\alpha f^2}} )$. In addition to $\varphi_0 =0$, the potential  also becomes zero at the points $\varphi_{\pm} = f \pm \frac{\omega}{\sqrt{\alpha}}$.
   
In this regime, the existence of a stationary bubble solution is rather transparent. The ``particle'' starts at the ``time'' $r=0$ with some initial position $\varphi(0)$. Notice, in order to tame the friction term, which becomes singular at $r=0$,  the initial velocity is fixed to be zero, $\mathrm{d}_r\varphi(r)|_{r=0} =0$. This leaves the initial position $\varphi(0)$ as the only adjustable initial condition. In order for the 
particle to reach $\varphi_0=0$ in asymptotic 
future $r=\infty$, the initial coordinate must satisfy $\varphi_- < \varphi(0) \leqslant \varphi_{\text{max}}$. The initial acceleration is then determined from, 
\begin{equation}
\mathrm{d}_r^2\varphi|_{r=0} = \varphi(0) \left( - \omega^2 + \alpha  (\varphi(0) - f)( 2 \varphi(0) - f)\right ) \, .
\end{equation}
It is clear that the solution with $\varphi(\infty) = 0$ can be obtained by continuously varying $\varphi(0)$. Indeed, taking $\varphi(0) = \varphi_-$ the particle always undershoots, since it inevitably loses energy due to a non-zero friction. On the contrary, by taking $\varphi(0)$ sufficiently close to $\varphi_{\text{max}}$, we can make the initial acceleration arbitrarily small, which will give the particle enough time not to move until the friction becomes negligible. In such a case, the particle would reach $\varphi_0=0$ with non-zero kinetic energy within a finite time. Obviously, between the two extremes there exists an initial $\varphi(0)$ which makes the friction just right for reaching $\varphi_0 =0$ for $r=\infty$.  \\
    
The study of the bubble solution simplifies in thin-wall approximation, which is given by,
\begin{equation} \label{TWALL}
\dfrac{\omega^2}{\alpha f^2} \, \ll \, 1 \, , ~ \, \text{or equivalently}, \, ~  \omega^2 \, \ll \, m^2 \, .
\end{equation}    
The above implies that the Goldstone frequency $\omega$ is much smaller than the mass gap $m$ in the symmetric vacuum. The quantum meaning of this regime shall become more transparent later. \\
       
In the above regime, the energy splitting between the maxima $\varphi_{\text{max}} \simeq f (1 + \frac{\omega^2}{\alpha f^2} )$ and $\varphi_0=0$ is approximately given by
\begin{equation}
V_{\text{max}} \simeq \dfrac{1}{2} \omega^2 f^2 \left( 1 + \dfrac{\omega^2}{\alpha f^2} \right) \, .
\end{equation}
In the thin-wall approximation, for a given $\omega$ the radius $R$ of the bubble can be found by extremizing the action
\begin{equation}
S = 4\pi \int\limits_0^{\infty} r^2 \, \mathrm{d}r \, \dfrac{1}{2} \left( \mathrm{d}_r \varphi(r) \right)^2 - V \left( \varphi(r) \right) \, ,
\end{equation}
evaluated on the bubble solution $\varphi(r)$. In the thin-wall regime, the particle spends the time interval $ 0 < r < R$ near the initial value close to the maximum $\varphi_{\text{max}}$. During this time, $\varphi(r)$ is essentially constant and the main contribution to the action comes from the potential energy at the maximum. This portion of the action is therefore given by $S_{\text{int}} \simeq  - \frac{4\pi}{3}V \left( \varphi_{\text{max}} \right) R^3 \simeq -\frac{2\pi}{3} \omega^2 \frac{m^2}{\alpha}R^3$. The  bulk of the transition from $\varphi_{\text{max}}$ to $\varphi_0=0$ is completed during the time interval $\Delta r = m^{-1}$, which sets the thickness of the bubble wall. Correspondingly, this part of the action is given by, $S_{\text{wall}} \simeq \frac{2\pi}{3}\frac{m^3}{\alpha} R^2$. Thus, the total action on thin-wall bubble solution is,
 \begin{equation}
 S_{\text{bubble}} = \dfrac{2\pi}{3}\frac{m^3}{\alpha} \left(R^2 
 -\dfrac{\omega^2}{m} R^3 \right) \, . 
 \end{equation}
 Extremizing this expression with respect to $R$, gives the following value of the radius of a stationary bubble, 
 \begin{equation} \label{thinwallR}
  R = \dfrac{2}{3}\frac{m}{\omega^2}
  \end{equation} 
   Notice that for $\omega \ll m$, we have,
  \begin{equation}
  R  \gg \frac{1}{m} \, .
  \end{equation} 
 Of course, this matches the fact that in the thin-wall approximation the bubble radius is much larger than the thickness of the wall, which is given by $m^{-1}$.
   
The solution $\varphi(r)$ obtained in this way, describes a stationary spherically-symmetric bubble in Minkowski space. The energy of such a bubble consists of the energy of the wall tension and the energy of the interior (both positive), 
    \begin{equation}
 E_{\text{bubble}} =  E_{\text{int}} + E_{\text{wall}} = \dfrac{\omega}{\alpha}\,  \frac{m^5}{\omega^5} \, \left (\frac{40\pi}{81} \right ) \, ,
 \end{equation} 
where, 
 \begin{equation}  
 E_{\text{int}} = \dfrac{2}{3} E_{\text{wall}} \, . 
\end{equation}
 
 \subsection{Quantum picture of classical stability} 
 
The wall of the bubble is predominantly made of the radial excitations of the $\varphi$ field of mass $m$,  whereas the interior is predominantly made of Goldstone excitations of frequency $\omega$. It is therefore convenient to write the two energies in terms of the occupation numbers of the corresponding quanta,      
\begin{equation}
 E_{\text{int}} = \omega N_G \, , ~~ \text{where}, \, ~~ N_G \equiv \dfrac{1}{\alpha} \dfrac{m^5}{\omega^5} \left(\dfrac{16\pi}{81}\right ) \, ,
\label{Eint}
\end{equation}
and 
  \begin{equation}
 E_{\text{wall}} =  m  N_{\varphi} \, , ~~ \text{where}, \, ~~ N_{\varphi} \equiv \dfrac{1}{\alpha} \, \dfrac{m^4}{\omega^4} \,
 \left(  \dfrac{8\pi}{27} \right ) \, .
\end{equation}
 We see that,  
  \begin{equation} \label{NGNP}
 \dfrac{N_G}{N_{\varphi}}
  = \frac{2}{3} \frac{m}{\omega} \gg 1 \, .
  \end{equation}     
   That is, the thin-wall bubble contains much higher occupation number 
   of the Goldstones than of the massive $\varphi$-quanta.
   
Thus, we arrive to the conclusion that a stationary bubble is obtained thanks to the excitations of the Goldstone modes. Let us try to understand this effect in the quantum language. In the thin wall regime (\ref{TWALL}), the Goldstone frequency $\omega$  in the bubble interior is much less than the mass of the quanta $m$ in the asymptotic vacuum outside of the bubble. Now, the interior energy (\ref{Eint}), which amounts to $2/3$ of the wall energy, is given by the Goldstone excitations.  
 
The bubble is stable because of two factors: 1) The fact that the Goldstone $SU(N)$ charge is conserved; and 2) the fact that the same amount of charge in the exterior vacuum would cost much higher energy. 
   
If the bubble would decay, the $SU(N)$ charge, stored in the Goldstone modes in the interior of the bubble, must be released into the asymptotic $SU(N)$-symmetric vacuum. But, this vacuum contains no gapless excitations and the lowest energy cost per particle is $m$. Carrying the entire charge in the form of massive particles would cost more energy than the energy of the bubble. This is the source of stability.

 \section{Closer look at Goldstones} 
 
 \subsection{Goldstones of internal symmetry}

 The spontaneous breaking of global 
 $SU(N)$ symmetry by the solution (\ref{ansatz}) results into 
 the appearance of Goldstone species localized in the bubble interior.  As already discussed, at large $N$, their number scales as $N_{\rm Gold} \simeq 4N$.  These Goldstone species 
 possess the tower of eigenmodes  $\theta^a_{\epsilon}$ of various  
 eigenfrequencies $\epsilon$.  In the limit of an infinitely
 large bubble, these eigenmodes reproduce the momentum 
modes of free Goldstone plane waves.  
 
 The exact form of 
 mode functions can be obtained by the standard mode analysis 
 of linearized fluctuations in the background
 of the classical solution.  This will not be presented here. 
 Instead, we shall focus on the resulting eigenmodes of two specific frequencies. 
  
  The modes $\theta_0^a$ corresponding to 
  $\epsilon =0$ have zero frequencies. These modes 
  account for the degeneracy of the bubble ``vacuum'' with respect 
  to broken $SU(N)$ transformations. In other words, 
  by acting on a bubble with a global $SU(N)$ transformation, 
  we excite a set of corresponding gapless Goldstone modes. 
  Of course, since these modes are gapless, their excitations cost no energy. This accounts for strict degeneracy of bubbles that are  
  related by global $SU(N)$ transformations. 
  Such bubbles form irreducible representations of 
  $SU(N)$ symmetry. 
   
  The second set of harmonics $\theta_{\omega}^a$ correspond  
   to eigenfrequencies $\epsilon = \omega$. 
These harmonics represent the actual constituents of the bubble. 
On the bubble solution, at least some of them have non-zero occupation numbers. When we act on a bubble
by an $SU(N)$ transformation, we change the occupation numbers 
of  $\theta_{\omega}^a$ modes without affecting the energy of the state.

To summarize, various Goldstone modes contribute into the portrait of the stationary bubble as follows. The energy of the interior of the bubble (\ref{Eint}) represents the energy of a state of Goldstone bosons $\theta_{\omega}^a $ of frequencies $\omega$ and a total mean occupation number, $N_G$. In the original ansatz (\ref{ansatz}), this entire occupation number was attributed to a single mode $a=1$. However, it can be arbitrarily redistributed among the Goldstones of different flavors due to $SU(N)$ symmetry of the theory. All the solutions obtained by the $SU(N)$ transformation are degenerate in energy. 
This degeneracy is accounted for by the presence of gapless Goldstone modes $\theta_0^a$.

The occupation numbers can be redistributed arbitrarily among the excited primary Goldstones of frequency $\omega$ subject to the constraint (we use large-$N$ approximation), 
\begin{equation} \label{ConstN}
 \sum\limits_a^{2N} n^a =  N_G \, . 
\end{equation} 
 Each sequence of such numbers $n^a$, represents a memory pattern, 
 \begin{equation} \label{pattern}
 \ket{Pattern} = \ket{n_{\omega}^1, n_{\omega}^2, \, \dotsc } \, ,
 \end{equation} 
 All of them amount to the classical solutions that are degenerate in energy. All such information patterns cost the energy given by (\ref{Eint}). That is, $E_{\text{int}}$ is the energy of the information pattern encoded in the $SU(N)$ charge via the excitations of Goldstones of frequency $\omega$. \\
 
 Now, it is important to understand the following distinction. Let us consider the two patterns, $\ket{n_{\omega}^1, n_{\omega}^2, \, \dotsc}$ and $\ket{n_{\omega}^{\prime 1}, n_{\omega}^{\prime 2}, \, \dotsc}$ for which the differences $\Delta n_{\omega}^{a} = | n_{\omega}^{a} -  n_{\omega}^{\prime a}|$ among some occupation numbers are large. 
 To be more precise, let us say that for some $a$-s, the 
 ratio $\frac{\Delta n_{\omega}^{a}}{N_G}$ is nonzero for 
 $N_G \rightarrow \infty$.  Then, such patterns are distinguishable in classical theory.
   
In addition to these, the set of all possible degenerate patterns contains subsets which have small differences. Such patterns are not distinguishable classically. Of course, they nevertheless contribute into the count of quantum micro-states. 
Correspondingly, the memory burden effect has a part that is classically observable and a part that can only be resolved in a quantum theory. 
 
Notice, as it is clear from (\ref{Eint}) in the thin-wall approximation  (\ref{TWALL}) we have, 
\begin{equation} 
N_G \sim \dfrac{1}{\alpha} \dfrac{m^5}{\omega^5}
\gg  \dfrac{1}{\alpha} \sim N_{\text{Gold}} \, .
 \end{equation}
Now remembering that the number of Goldstone flavors scales as $N_{\text{Gold}} \sim N$, and that  unitarity puts the upper bound $N \sim 1/\alpha$, we get that for a thin-wall bubble, 
 \begin{equation} 
N_G   \gg  N_{\text{Gold}} \, .
 \end{equation}
That is, in thin-wall regime the occupation number of the Goldstone modes is much larger than the number of their species. Due to this, the main stabilizing force is the classical part of the memory burden. For smaller bubbles, the situation changes dramatically and the quantum part of the memory burden becomes as important as the classical one. \\ 
   
If the bubble decays, the information stored in the memory pattern (\ref{pattern}) must be released in the form of asymptotic quanta of mass $m$. Since the $SU(N)$ charges are conserved, the minimal energy cost of such an asymptotic state is 
 \begin{equation}
 E_{\text{pattern}} = m N_G \, ,
 \end{equation}  
  where $N_G$ is given by (\ref{Eint}). 
  This can be rewritten as, 
\begin{equation} 
 E_{\text{pattern}} \, = \, mN_G \, = \,  m \,\dfrac{1}{\alpha}  \,\dfrac{m^5}{\omega^5} \left (\dfrac{16\pi}{81} \right ) \, =\,  \dfrac{2m}{5\omega}\, E_{\text{bubble}} \, .
 \end{equation}
  Thus, the energy cost of the information pattern 
  released in case of a  bubble decay would 
 exceed the energy of the bubble by a factor  $2m/(5\omega)$. This is of course impossible. It is therefore energetically more favourable to store this amount of charge in the form of the bubble, rather than to release it in the form of the asymptotic quanta. 
  \\

 One may wonder whether
instead of free quanta the Goldstone charge can be released in the form of the smaller bound states with more
effective mass-to-charge ratio.   
However,  this does not appear plausible.  
Due to the weak coupling $\alpha$ it takes 
at least $\sim \frac{1}{\alpha}$ particles to make up a  
bound state of size $R \sim 1/m$.  This is the smallest 
possible size for a bound state due to the energy scale in the problem.  
Such a bound state 
will have energy $\sim m/\alpha$ and a charge 
capacity $\sim 1/\alpha$.  In fact, it would represent a smallest size vacuum bubble. 
 It is clear that the charge of the large bubble 
cannot be released in the form of such objects. 
Basically, a bubble cannot decay into smaller bubbles. 
The stationary bubble is energetically the optimal configuration for the given charge, at least   
among the spherically symmetric ones. This is also confirmed by our numerical analysis.

Although we did not provide a proof, it is rather natural that for a bubble with zero angular momentum, the departure from the spherical symmetry cannot lower the energy, since there is no parameter 
in the theory that would set the measure of asymmetry for  the lowest energy configuration, which must exist due to impossibility 
of the decay into the free quanta.  

The important thing is that the bubble
is an existence proof of the bound state that stores the $SU(N)$ 
charge in a more energy-efficient way than the collection of 
the asymptotic quanta.  Thus, there exist 
bound states stabilized by the memory burden effect.

In this sense, the stabilized bubbles can be considered as sort of non-topological solitons, or $Q$-balls \cite{Lee:1991ax, Coleman:1985ki}. There exists a large literature on the subject. In particular, $Q$-balls with arbitrarily small classical charges were analyzed in \cite{Kusenko:1997ad}. The physical effect, such as catalysis of proton decay, due to symmetry breaking inside the $Q$-ball interior were also studied \cite{Dvali:1997qv}. In this sense, it can be said that the present paper also sheds a very different light on the mechanism of $Q$-ball stability, by interpreting it as a sort of memory burden effect. 
  
At the same time, the vacuum bubbles considered here are qualitatively different from the previously studied cases of $Q$-balls.  First of all, bubbles carry a large micro-state entropy which can saturate the unitarity bound. Due to this, the 
information encoded in the Goldstone modes that stabilizes 
the bubble, can be intrinsically quantum.  
  
This form of stabilization takes place in thick-wall bubbles, for which $\omega \sim m \sim 1/R$, where the quantum contribution into the memory burden becomes as important as the classical one. This is linked with the fact that such bubbles saturate the entropy bounds (\ref{Area}) and (\ref{alphaB}).  As saturons,  they satisfy $N_G \sim N_{\text{Gold}} \sim N \sim 1/\alpha$. 
 
\subsection{Goldstones of broken Poincar\'{e} symmetry}
 
Let us now focus on the Goldstone bosons of broken Poincar\'{e} symmetry.  They consist of broken space and time translations. Of course, correspondingly, the Lorentz boosts are also broken. Let us consider a thin-wall bubble first. \\
  
The different constituents of the bubble share the breaking of space-time symmetries as follows. The spontaneous breaking of space translations comes predominantly from the bubble walls. The main weight of breaking is carried by the 
$\varphi$-quanta of occupation number $N_{\varphi}$, each contributing $m$ in the order parameter. Therefore, up to order-one numerical factors, the couplings of the space-translation Goldstone bosons are, 
 \begin{equation}
     G_{\text{Gold}}^{(\text{s})} =  \dfrac{R}{mN_{\varphi}}  \,.    
  \end{equation}
In contrast with space translations, the dominant contribution into the breaking of time translation symmetry comes from the interior of the bubble. This is due to non-zero frequency $\omega$ of Goldstone modes of broken $SU(N)$-symmetry. Their total occupation number is $N_G$. Each occupied quantum contributes $\omega$ into the breaking of time-translations. The coupling of the time translation Goldstone mode therefore is, 
\begin{equation}
     G_{\text{Gold}}^{(\text{t})} =  \dfrac{R}{\omega N_G}  \,.    
  \end{equation}
From the relation (\ref{NGNP}) between the occupation numbers we have, 
 \begin{equation} \label{NumbersGP}
 \omega N_G \sim  m N_{\varphi} \, .   
  \end{equation}
It is therefore clear that the couplings of Goldstones of broken space and time translations are of the same order,   
 \begin{equation} \label{Gst}
     G_{\text{Gold}}^{(\text{s})} \sim G_{\text{Gold}}^{(\text{t})}  \sim 
       \frac{R}{\omega N_G} \, .    
  \end{equation}
This is not surprising, since the bubble walls that break space translations are stabilized due to breaking of time translation 
by its interior. \\
   
The relation (\ref{NumbersGP}), and correspondingly (\ref{Gst}), holds for an arbitrary classically-stable bubble. This includes the thick-wall regime. Therefore, the coupling of a generic Goldstone of broken Poincar\'{e} symmetry can be expressed in terms of the occupation number of $SU(N)$ Goldstones as, 
\begin{equation} \label{GPoincare}
     G_{\text{Gold}}^{(\text{P})} \sim \dfrac{R}{\omega}\dfrac{1}{N_G} 
 \sim \sqrt{\dfrac{R^3}{m}} \dfrac{1}{N_G}  \,,     
  \end{equation}
where we took into account the relation (\ref{thinwallR}) between $\omega$, $m$ and $R$. 

Now, the dimensionless effective coupling of the Poincar\'{e} Goldstone, $\alpha_{\text{Gold}}^{(\text{P})}$, evaluated at the energy scale $1/R$ (corresponding to the size of bubble) is, 
\begin{equation} \label{AlphaPoincare}
 \alpha_{\text{Gold}}^{(\text{P})} \equiv 
 G_{\text{Gold}}^{(\text{P})} \dfrac{1}{R^2} \, = \, \dfrac{G_{\text{Gold}}^{(\text{P})}}{{\mathcal 
 Area}} \, = \, \dfrac{1}{\sqrt{mR}}\dfrac{1}{N_G}\, = 
 \dfrac{\omega}{m}\dfrac{1}{N_G} \,,    
  \end{equation}
  where, as before, ${\mathcal Area} \sim R^2$ is the area of the bubble. 
  
Using (\ref{GPoincare}) and  (\ref{AlphaPoincare}) we can present the entropy bounds (\ref{Area}) and (\ref{alphaB}) for a bubble in terms of the occupation number of the 
$SU(N)$ Goldstone modes, 
 \begin{equation}  \label{SMAXNG}
   S_{\text{max}} \, \sim \,  \dfrac{1}{\alpha_{\text{Gold}}^{(\text{P})}} \sim 
    \dfrac{{\mathcal Area}}{G_{\text{Gold}}^{(\text{P})}} \,  
 \sim \,  N_G  \, \sqrt{mR} \, \sim \,  N_G\, \dfrac{m}{\omega} \,.     
  \end{equation}
The above represents the maximal entropy that can be attained by a bubble in an unitary theory. As we shall see, only the thick-wall bubbles are able to saturate this bound.

 \section{Spectrum of bubbles} 
 
  Our previous analysis shows that bubbles represent the bound-states of Goldstone bosons, with occupation number $N_G$. The energy spectrum of bubbles can be expressed as a function of $N_G$. For thin-wall bubbles we have,   
  \begin{equation}
 E_{N_G}  \, = \, \frac{5}{2} \, \omega \, N_G\, = \,
  \dfrac{5}{\sqrt{6}}\, \sqrt{\dfrac{m}{R}} \, N_G \, .
 \end{equation} 
 Up to an order-one  coefficient, the same relation holds for the thick-wall bubbles for which $m\sim 1/R$. As we shall see later, only thick-wall bubbles can saturate the unitarity bound on entropy. This happens in the regime, $N_G \sim N\sim 1/\alpha$. The energy of a saturon bubble is thus given by, 
   \begin{equation} \label{Esaturon1}
 E_{N} \,  \sim \,  \frac{N}{R} \, \sim \, \frac{1}{\alpha}\, \frac{1}{R} \, \sim \, 
   \frac{m}{\alpha} \,.
 \end{equation} 
 Thus, a saturon represents a bound state of $N_G \sim N$ Goldstone bosons and of the same order $N_{\varphi} \sim N$ radial modes.

 \section{Entropy of a bubble}
 
 Let us estimate the entropy of a bubble. The number of degenerate micro-states is given by all possible sequences (\ref{pattern}) satisfying the constraint (\ref{ConstN}). Using Stirling approximation, and working at leading order in large $N_G$ and $N$, this number can be written as  
 \begin{equation} \label{NstB}
n_{\text{st}} \sim \left( 1+\dfrac{2N}{N_G} \right)^{N_G} \left( 1+\dfrac{N_G}{2N} \right)^{2N} \,.
\end{equation}
  As explained in \cite{Dvali:2019jjw, Dvali:2019ulr, Dvali:2020wqi}, the above expression for the number of degenerate micro-states is common for solitons that spontaneously break global symmetry. It can be understood in the following two complementary languages. \\
  
  {\bf Internal view}:
   
From the point of view of the world-volume theory of the soliton, the degeneracy (\ref{NstB}) accounts for the degeneracy of the Goldstone vacuum. An observer located in the interior of the soliton (in the present case, a bubble), sees a spontaneous breaking of global $SU(N)$ symmetry with the resulting gapless Goldstone modes. Just as any other Goldstone vacuum, this vacuum is degenerate. However, since the bubble has a finite size, the number of 
independent orthogonal vacuum states is finite rather than infinite. This degeneracy becomes infinite in the limit of a bubble of an infinite extent. In this limit, the bubble 
vacuum becomes an ordinary Goldstone vacuum 
of four-dimensional theory. 

To summarize, according to internal view, the different micro-states counted in (\ref{NstB}) are Goldstone vacua that are obtained from one another by the action of $SU(N)$ transformations. \\
  
 {\bf External view}:
 
From the point of view of an external observer, the same degeneracy is explained differently. This observer lives in the asymptotic vacuum for which $SU(N)$ is a good symmetry. Correspondingly, all the states can be classified according to representations of the $SU(N)$ group. \\
   
Thus, the external observer sees the bubble as an object transforming under one of such representations. 
 This representation is exponentially large. This is because the bubble represents a bound state of large number of quanta each transforming under the adjoint representation of $SU(N)$. Let the total occupation number of quanta 
be $N_T$. Then, the bubble transforms as a tensor product of $N_T$ adjoints. The wave function of the bubble can be written as a tensor, 
 \begin{equation} \label{BWF}
   {\mathcal B}_{\alpha_1,\alpha_2 , \, \dotsc \, , \alpha_{N_T}}^ 
   {\beta_1,\beta_2 , \, \dotsc \, , \beta_{N_T}} \, ,
 \end{equation} 
 which is totally symmetric under lower and upper indices and has a zero trace with respect to each conjugated pair. To the leading order, the dimensionality of such a tensor is given by the square of the binomial coefficient. Using the Stirling approximation it can be written as,  
  \begin{equation} \label{NSTNT}
n_{\text{st}}  \sim 
\left( 1+\dfrac{2N}{N_T} \right)^{N_T} \left( 1+\dfrac{N_T}{2N} \right)^{2N} \,.
\end{equation}
 Now, our previous analysis shows that $N_T$ is well approximated by $N_G$. Indeed, as it is clear from (\ref{NGNP}), for the thin-wall bubble (\ref{TWALL}), the occupation number of Goldstones $N_G$ dominates over the number of other constituents. So, in this regime we have $N_T \simeq N_G$. For the thick-wall regime, the occupation number of the non-Goldstone quanta, $N_{\varphi}$, becomes comparable to $N_G$. Thus, without loss of generality we can write $N_T \sim N_G$.  It is then clear that the expression (\ref{NSTNT}) gives the same count of the micro-states as (\ref{NstB}). \\ 
 
The corresponding micro-state entropy is given by, 
  \begin{equation} \label{SLambda}
S = \ln (n_{\text{st}}) \simeq  2N \, \ln \left[
 \left( 1+ \lambda \right)^{\frac{1}{\lambda}} \left( 
 1+\dfrac{1}{\lambda} \right) \right] \,.
\end{equation}
where we have introduced a notation $\lambda \equiv 
2N/N_T \sim 2N/N_G$. This quantity represents the ratio of the number of primary Goldstone species to their occupation number in the bubble state. This ratio decides how far is the entropy of the system from saturating the bounds imposed by  unitarity. We shall study this question next.

\section{Saturation} 

We now wish to investigate, in which regime a bubble saturates the bounds on entropy (\ref{Area}) and (\ref{alphaB}). Naively, it may appear that the entropy (\ref{SLambda}) can be made arbitrarily large, at the expense of increasing $N$. However, this is not the case, since $N$ and $\alpha$ are ``locked'' in the unitarity constraint (\ref{Ubound}). Due to this, $N$ reaches its maximum when the 't Hooft coupling of the theory is taken at its unitarity upper bound, 
 \begin{equation} \label{Ccrit} 
   N \alpha  \sim 1 \,. 
\end{equation} 
As explained in \cite{Dvali:2020wqi}, any further growth of the 't Hooft coupling beyond this point, must lead to the change of the regime. This is signalled both by the breakdown of the loop expansion, as well as, by saturation of unitarity by various scattering amplitudes. \\
 
Now, since the number of Goldstone species $N_{\text{Gold}}$ is of order $N$, the condition (\ref{Ccrit}) implies,
   \begin{equation} \label{NSPcrit} 
   N_{\text{Gold}} \, \sim \, N\,  \sim \, \dfrac{1}{\alpha} \,. 
\end{equation}
The entropy of the bubble must be evaluated subject to this constraint. In what follows, we shall perform this evaluation in two different regimes.

\subsection{Entropy of thin-wall bubbles}

Let us start by considering the thin-wall regime. From (\ref{Eint}) it is clear that for a thin-wall bubble we have,  
 \begin{equation}
   N_G \, \gg \, \frac{1}{\alpha} \,.   
\end{equation} 
Taking into account the expressions (\ref{Ccrit}) and (\ref{NSPcrit}), it becomes obvious that in a thin-wall bubble we have $\lambda \ll 1$. That is, the thin-wall bubble satisfies,  
  \begin{equation}
   N_G  \gg N_{\text{Gold}}  \sim N \, \sim \dfrac{1}{\alpha} \,. 
\end{equation} 
Therefore, in this regime the expression (\ref{SLambda}) becomes, 
 \begin{equation} \label{Sthinwall}
S \,   \simeq \, 2N \, \ln \left (\dfrac{\mathrm{e}}{\lambda} \right)\,  
 \sim  \dfrac{1}{\alpha} \, \ln \left(\dfrac{m^{10}}{\omega^{10}} \right ) \,.
\end{equation}
In the last term we took into account (\ref{Ccrit}), expressed $N_G$ as function of $\alpha, \omega, m$ via (\ref{Eint}) and ignored order-one numerical factors ($8\mathrm{e}\pi/81 \sim 1$) in the logarithm.  \\ 
   
The actual entropy of the bubble (\ref{Sthinwall})
 is much smaller than the upper bound (\ref{Area}) applied to the same bubble. Indeed, expressing $N_G$ via (\ref{Eint}), the maximal entropy (\ref{SMAXNG}) of a thin-wall bubble, permitted by the Poincar\'{e} Goldstone is, 
 \begin{equation}
S_{\text{max}}  
 \sim \dfrac{1}{\alpha} \dfrac{m^6}{\omega^6} \,.
\end{equation}
  Obviously, for a thin-wall bubble ($m/\omega \gg 1$) this is much larger than the actual entropy (\ref{Sthinwall}), which scales only logarithmically with $m/\omega$. Thus, a thin-wall bubble represents an under-saturated state in terms of its entropy capacity. \\
      
The same conclusion can be reached by examining the entropy bounds posed by $SU(N)$ Goldstones. One can see this, by repeating the logic of \cite{Dvali:2020wqi}. Let us consider an effective dimensionless coupling of $SU(N)$ Goldstones, evaluated at the scale $1/R$. It is equal to, 
  \begin{equation}
\alpha_{\text{Gold}}^{(SU(N))} \, = \, \dfrac{1}{(fR)^2} \, \sim  \,  \alpha \dfrac{\omega^4}{m^4} \,. 
\end{equation} 
This coupling controls the scattering processes of Goldstones with momentum-transfer $\sim 1/R$. Let us now put ourselves in the position of an observer living in the effective low energy theory of Goldstones with the cutoff $\Lambda_{\text{cutoff}} \sim 1/R$. 
  
For the consistency of this limited sector, the observer only needs to make sure that the effective theory is unitary up to the scale $\sim 1/R$.  With this minimal requirement, the observer has the ``luxury'' of introducing as many as $N_{\text{Gold}} \sim 1/\alpha_{\text{Gold}}^{(SU(N))}$ species of Goldstones.  
Such a large diversity of Goldstone flavors would endow the bubble with the micro-state entropy, 
 \begin{equation}
S_{\text{max}}  
 \sim \dfrac{1}{\alpha} \, \dfrac{m^4}{\omega^4} \,.
\label{SPareaSUN}
\end{equation} 
Therefore, a low energy observer would conclude that the bubble of size 
$R$ can attain the above entropy. However, this would certainly run the 
full theory into a problem which is beyond the knowledge of the low energy observer.
  
Indeed, since the number of Goldstone species inhabiting the bubble is set by $N$, attaining (\ref{SPareaSUN}) would require an increase of $N$ up to $N \, \sim \, \frac{1}{\alpha} \frac{m^4}{\omega^4}$. However, this would  violate the unitarity constraint (\ref{Ubound}) of the full theory. This 
of course  is not possible.
   
The unitarity constraint (\ref{Ubound}) precludes the introduction of Goldstone species with the number above (\ref{NSPcrit}). This limits the actual entropy capacity of the bubble by (\ref{Sthinwall}). This limiting entropy is much smaller than (\ref{SPareaSUN}) which is mistakenly presumed by the effective low energy theory with the cutoff $\Lambda_{\text{cutoff}} \sim 1/R$.
    
 We thus arrive to the following conclusion. In the thin-wall regime, the entropy of the bubble (\ref{Sthinwall}) is far from saturating the bounds (\ref{Area}) and (\ref{alphaB}).  
 
 \subsection{Entropy of thick-wall bubbles}
 
  The situation changes drastically for thick-wall  bubbles, for which $R \sim 1/m$. As already explained, as long as condition (\ref{MinC}) is satisfied, (or equivalently, if $\omega < m $), the existence of a classical solution of a stationary bubble, described by (\ref{ansatz}) and (\ref{BCofansatz}) is guaranteed. For definiteness, we shall assume that the condition (\ref{MinC}) is satisfied and we shall focus on the case $\omega \sim m$.
     
Obviously, for such a bubble, all three scales are of the same order,
 \begin{equation} \label{OMR}
   \omega \sim m \sim \dfrac{1}{R}\,. 
  \end{equation} 
  Correspondingly, we have, 
  \begin{equation} \label{NGsmall}
  N_G \, \sim \, \dfrac{1}{\alpha}\,. 
  \end{equation} 
  Taking into account (\ref{Ccrit}), from (\ref{SLambda}) it follows that the entropy is,  
    \begin{equation} \label{SsmallB}
S \sim \dfrac{1}{\alpha} \,.
\end{equation}
Thus, the entropy saturates the unitarity bound (\ref{alphaB}). 

The area-law bound (\ref{Area}) is also saturated. Indeed, since the size of the bubble satisfies (\ref{OMR}), the coupling of the Poincar\'{e} Goldstone (\ref{GPoincare}) becomes, 
  \begin{equation}
  G_{\text{Gold}}^{(\text{P})} \, \sim \, \dfrac{1}{N} \, {\mathcal Area} \, \sim \,  \alpha \,  {\mathcal Area}  \,.
  \end{equation} 
 It is clear that the entropy of the bubble (\ref{SsmallB}) saturates the bound (\ref{Area}).  
   
It is also obvious that for a saturated bubble, the couplings of the Poincar\'{e} and internal Goldstones are equal. So are their decay constants. We shall therefore drop the subscripts and denote these quantities universally by $G_{\text{Gold}}$ and $f$ respectively. Also, from now on, we shall denote the dimensionless coupling of Goldstones by $\alpha_G$.  

   For both Poincar\'{e} and internal Goldstones, the expressions for the couplings and the decay constants are given by (\ref{Gdecayf}), which we shall rewrite as, 
  \begin{equation} 
  G_{\text{Gold}}\,  =\,  f^{-2}\,  = \, \dfrac{1}{N} \, {\mathcal Area} \,   \sim \, \alpha  \,  {\mathcal Area}  \,.
  \end{equation} 
  Similarly, the dimensionless couplings $\alpha_G$ of both types of Goldstones, evaluated at the scale $1/R$, are equal to the fundamental coupling $\alpha$. That is, in the thick-wall regime we have
  \begin{equation} \label{AAA1}
  \alpha_G = \dfrac{G_{\text{Gold}}}{{\mathcal Area}} 
= \dfrac{1}{(fR)^2} = \alpha \,. 
  \end{equation} 
    Correspondingly, the bounds (\ref{Area}) and (\ref{alphaB}) are universally saturated in terms of all couplings and all the decay  constants of the theory. \\

 Interestingly, as already noticed in \cite{Dvali:2020wqi}, 
the entropy of the saturon bubble obeys, 
 \begin{equation} \label{BEK}
  S \, \sim \, E_{\rm bubble} \, R \,. 
  \end{equation} 
  This is easy to see from (\ref{Esaturon1}), 
  (\ref{SsmallB}) and (\ref{OMR}). 
 The above expression agrees with the bound on entropy 
 suggested by Bekenstein \cite{Bekenstein:1980jp}, despite the fact that {\it a priori}  
 the latter bound has no information about the coupling 
 of the system.  However, for generic bubbles the bounds (\ref{Area}) and (\ref{alphaB}) appear more stringent, as
 they exclude an excessive entropy of a bubble even 
 if the Bekenstein bound is formally satisfied.    
    \\

In summary, we observe that the thick-wall bubbles saturate the entropy bounds (\ref{Area}) and (\ref{alphaB}), when the theory saturates unitarity. That is, a thick-wall bubble becomes a saturon when the 't Hooft coupling is critical (\ref{Ubound}). At the same time, the thin-wall bubbles are under-saturated. That is, their entropy is much less than what would be permitted by the effective low-energy theory of Goldstones, with the cutoff $\sim 1/R$. The thin-wall bubble cannot achieve such entropy, without invalidating the full theory. The results are summarized in table \ref{tab_entropies}. \footnote{~Interestingly, the same conclusion was reached in \cite{Dvali:2020wqi} for thin and thick wall bubbles that 
were not stabilized by memory burden of the Goldstone charge but instead were assumed to freely oscillate (and eventually decay).}  
\\
 
  \begin{table}[H]
\centering
\begin{tabular}{c|c|c}
\multicolumn{1}{c|}{\diagbox[innerwidth=3.1cm]{Bubble}{Entropy}} & Bound on $S$ & Actual $S$ \\
\hline
thin-wall, $\frac{m}{\omega} \gg 1$ & $S\simeq \frac{1}{\alpha} \frac{m^6}{\omega^6} $ & $S\simeq \frac{1}{\alpha} \ln \left(\frac{m^{10}}{\omega^{10}} \right )$ \\
thick-wall, $\frac{m}{\omega} \sim  1$ & $S \sim \frac{1}{\alpha}$ & $S \sim \frac{1}{\alpha}$ \\
\end{tabular}
\caption{Entropy bounds and the actual entropies for thin-wall and thick-wall bubbles. It is clear that thin-wall bubbles are under-saturated. The saturons originate from thick-wall bubbles.}
\label{tab_entropies}
\end{table} 
  
The above reveals a deep connection between the saturation of the system and the nature of the stabilizing memory burden effect. For under-saturated bubbles, the occupation number of Goldstones $N_G$ is much higher than the number of the Goldstone species $N_{\text{Gold}} \sim N$. Due to this, the memory burden that stabilizes the under-saturated bubble is classical. In contrast, a saturated bubble can be stabilized by the memory burden that is fully quantum in its nature. The physical meaning of this statement shall be explained below.

  \section{Stabilization by quantum memory burden}
  
   We have seen that when the collective coupling of the theory hits the unitarity bound (\ref{Ccrit}), the thin-wall bubble becomes a saturon. In this regime we are dealing with an interesting phenomenon. Despite the fact that the stationary bubble solution is well described classically, the memory burden stabilizing it can be fully quantum.
  
    In order to see this, let us start with the ansatz (\ref{ansatz}). As we already discussed, in the quantum language this classical solution corresponds to the state in which a single Goldstone mode, say $\theta^1$, is macroscopically occupied. The occupation number is given by (\ref{NGsmall}). Since our analysis becomes only better at weak coupling $\alpha$, the occupation number $N_G \sim N$ can be made arbitrarily large by taking $\alpha$ small. Such a state can be treated classically up to corrections $\sim 1/N$. \\

Now, we should remember that in the current regime the number of Goldstone species $N_{\text{Gold}}$ is of the same order as the total occupation number $N_G$. Both of these numbers are of order $N$. In the classical bubble, corresponding to the ansatz (\ref{ansatz}), the entire occupation number is ``credited'' to a single Goldstone flavor, $\theta^1$. The corresponding memory pattern (\ref{pattern}) has the form $n_{\omega}^a =\delta^{a1} N_G$. \\
   
However, as we already discussed, due to $SU(N)$ symmetry, there exist bubbles in which the number $N_G$ is distributed among various Goldstone flavors, subject to the constraint (\ref{ConstN}). Such states are related to each other by $SU(N)$ symmetry transformations. Due to this, they are all strictly degenerate in energy. \\
 
Among the patterns (\ref{pattern}), we wish to pay special attention to the ones in which the number is distributed approximately equally among all of the Goldstone flavors. Since for the saturated bubble $N_G \sim N_{\text{Gold}} \sim N$, in such states all occupation numbers are small.
The corresponding memory patterns (\ref{pattern}) have $n_{\omega}^a \sim 1$ for all $a$. Since the individual occupation numbers are small, we can say that the Goldstone modes are in intrinsically-quantum states.  
             
By power of symmetry, it is obvious that the corresponding memory burden has the same stabilizing effect as the one $n_{\omega}^a =\delta^{a1} N_G$, which is described by the classical solution (\ref{ansatz}). Thus, although the states (\ref{pattern}) with $n_{\omega}^a \sim 1$ are quantum, they possess the same stabilizing power as the ``classical'' state $n_{\omega}^a =\delta^{a1}N_G$ to which they are related by the $SU(N)$ symmetry. In other words, in such states, the stabilization happens due to a quantum memory burden. \\
   
We thus observe that for saturon bubbles, the stabilization by the memory pattern can be fully quantum. Such a bubble is similar to a black hole.

  \section{Information horizon} 
  
 As, already pointed out in \cite{Dvali:2020wqi}, an universal property of the saturons is that in semi-classical limit, they possess a strict information horizon. That is, the information stored in the interior of the saturon cannot be extracted in any form. The general physical reason is that in this limit the memory modes, that carry quantum information, decouple. Correspondingly, the information stored in these modes becomes permanently hidden.  \\
  
 In the present construction we can demonstrate this feature explicitly. The memory modes are represented by the Goldstone modes of spontaneously broken $SU(N)$ symmetry. Their interaction strength is suppressed by their decay constant $f$. The effective coupling of a Goldstone mode of frequency $\epsilon$ is, 
 \begin{equation} \label{AAA2}
  \alpha_G = \dfrac{\epsilon^2}{f^2} \,. 
  \end{equation} 
In the semi-classical limit this coupling goes to zero for any finite $\epsilon$. \\
   
In order to see this, we must specify the correct semi-classical limit. This is the limit in which the classical bubble solution with finite radius $R$ and frequency $\omega$ experiences no back reaction from quantum fluctuations. Such a limit is uniquely given by 
 \begin{equation} \label{Limit}
   \alpha \rightarrow 0\,,~ \, R= \text{finite}\,, ~ \omega =
   \text{finite} \,, ~\,  \alpha N = \text{finite} \,.     
 \end{equation}
Notice, this implies that the mass $m$ stays finite, whereas the Goldstone decay constant $f$ becomes infinite. Therefore, at the same time, in the above limit we have 
 \begin{equation} \label{AAA3}
 f \rightarrow \infty\,, ~ \alpha_G  \rightarrow 0\,, 
  \end{equation} 
for any finite $\epsilon$. Thus, the coupling of a Goldstone mode of arbitrarily high frequency vanishes. This applies to the Goldstones of frequencies higher than the mass $m$ of the quanta outside of the bubble.  Correspondingly, in the zero back reaction limit, there is no way of transmitting any information from the interior of the bubble to the exterior. \\
  
Even if the energy of a Goldstone perturbation is much higher than the mass gap in the asymptotic vacuum $r \gg R$, it is impossible for such a mode to communicate with the ones propagating in the outside vacuum. The bubble literally possesses an information horizon, analogous to that of a black hole.  \\
   
Another side of the story is the existence of the Goldstone horizon which manifests itself already at finite $f$. Goldstone waves of frequencies $\epsilon \ll m$ cannot propagate outside the bubble, even though at finite $f$ the coupling among the modes is finite. Here we can distinguish two cases.  \\
    
The first case is when the energy of an internal perturbation is less than $m$. In such a case propagation is simply impossible due to the finite energy gap.  \\
   
In the second case, although the frequency $\epsilon$ of the Goldstone perturbation is smaller than $m$, its energy can exceed the mass gap at the expense of the occupation number of Goldstone quanta in the perturbation. In the classical language, the amplitude of the Goldstone wave may be sufficiently high to reach the energy $m$, even if the frequency is very low. 
   
Even in such a case, the propagation is highly suppressed. The way to understand this is to think of a process as the transition from the initial state of high occupation number $n_{\epsilon}$ of soft quanta of frequency $\epsilon \ll m$, into a state of few quanta of frequencies $m$ or higher. For example, at the threshold, number $n_{\epsilon} = m/\epsilon$ of  Goldstones of frequencies $\epsilon$  can transit into a single exterior particle of mass $m$. Such a process is exponentially suppressed by a factor $\mathrm{e}^{-n_{\epsilon}}$ \cite{Dvali:2020wqi}. This is a general feature of the transition processes between high and low occupation numbers.

 \section{Hawking evaporation} 
 
  Despite their classical stability, the saturated vacuum bubbles can decay through quantum processes. As it is the case for generic saturons \cite{Dvali:2020wqi}, this decay is strikingly similar to Hawking evaporation of a black hole. Just like the case of a black hole, the emission has a thermal-like spectrum with a characteristic temperature $T$ given by the inverse radius of the bubble, 
  \begin{equation} \label{Temperature} 
      T \sim \dfrac{1}{R} \,.
 \end{equation}      
  In the semi-classical limit (\ref{Limit}), the quantum information carried by the radiated quanta is not resolvable. In full quantum theory, the resolution time is given by (\ref{Volume}) which is very similar to Page's time for a black hole. \\
    
In case of a black hole, the asymptotic theory is gapless, as it contains a massless field in the form of the graviton. Therefore, in order to make the connection between the saturated bubbles and black holes maximally transparent, we shall introduce some additional massless fields that can propagate in the asymptotic vacuum. \\
  
For definiteness, let us introduce a scalar $\xi_{\alpha}$ transforming in the fundamental representation of the $SU(N)$ group.  We assume that the Lagrangian is completed 
to a most general $SU(N)$-invariant 
renormalizable theory of  $\phi_{\alpha}^{\beta}$ and
$\xi_{\alpha}$. The only restriction we put is that in 
$SU(N)$-invariant vacuum $\xi_{\alpha}$ is massless. 

For making our point, it suffices to consider the following 
terms,  
 \begin{equation}
L_{\xi} = (\partial_{\mu}  \xi^{\dagger \alpha} )
( \partial^{\mu} \xi_{\alpha} ) \, - \, 
  \alpha_{\xi} \xi^{\dagger \alpha}  (\phi^2)_{\alpha}^{\beta}  \xi_{\beta} \,, 
  \end{equation} 
  where $\alpha_{\xi} > 0$ is a coupling constant, which is subject to the unitarity constraint, 
 \begin{equation}
   N \alpha_{\xi} \lesssim 1\,, 
   \end{equation}
  similar to the bound (\ref{Ubound})  obeyed by the coupling $\alpha$.   We shall choose $\alpha_{\xi}$ to be near the saturation point of the above bound, 
  $\alpha_{\xi} \sim 1/N$. \\  
      
  Now, as said, in the vacuum with unbroken $SU(N)$ symmetry, the field $\xi$ is massless. In the $SU(N-1)\times U(1)$-symmetric vacuum, its components get positive mass terms of order $m_{\xi}^2 \sim \alpha_{\xi} f^2 \sim m^2$. 
  Let us now consider the $\xi$-field in the background of a 
 stationary vacuum bubble of radius $R$.  
  
 In the asymptotic vacuum outside the bubble, $r \gg R$, the $\xi$-field propagates as massless. At the same time, in the interior of the bubble it gets a positive mass term. Thus, effectively, $\xi$ sees the bubble as a potential barrier. \\
      
For a thin-wall bubble (\ref{TWALL}), due to the positive mass$^2$ inside the bubble, the wavefunctions of the $\xi$ modes of momenta $q \ll m$ are exponentially suppressed (``screened'') in the bubble center. The screening factor is $\sim \mathrm{e}^{-mR}$. \\
   
For the saturated bubble, the above suppression is not significant. Such bubbles are in thick-wall regime and satisfy (\ref{OMR}). Due to this, the radius of the bubble $R$ is of the same order as the screening depth, which is given by $1/m$. In addition, $\omega \sim m$. Due to this, the $\xi$ modes in general have order-one overlap with the interior region of the saturated bubble.  \\ 
   
The presence of a massless field $\xi$ in the theory does not affect the classical stability of the bubble. However, it opens up a new quantum decay channel. In the semi-classical approximation, the decay rate can be derived by quantization of $\xi$ in the background of the classical bubble solution (\ref{ansatz}). However, since the saturation takes place in a thick-wall regime, where the solution is not known analytically,  such analysis will be purely qualitative and shall not be displayed here. \\
      
Instead, we shall take a shortcut directly into a full quantum theory and try to exploit the power of Goldstone universality. Namely, we shall use the fact that the couplings of Goldstones are governed by the symmetry breaking order parameter. Also, as already explained, most of the quantum information of the bubble, is carried by the Goldstone modes. Due to this, the decays of the bubble through the Goldstone processes are the most interesting from the perspective of the retrieval of quantum information by an outside observer. \\
  
The couplings between the Goldstone modes and the modes of $\xi$-quanta, are uniquely defined by the Goldstone theorem and are controlled by a single scale. In the broken-symmetry vacuum, in the language of a full (3+1)-dimensional theory, this coupling is local and has the form, 
  \begin{equation}
    i \partial_{\mu} \theta^{a} \, \left( \partial^{\mu}
     \xi^{\dagger}T^{a}  \xi  \, - \,
    \xi^{\dagger} T^{a} \partial^{\mu}\xi  \right )\,   + \, ... \,,
\end{equation}     
 where the dimensionless phases are related with the canonically normalized Goldstone fields 
 ($a^{a}$) through $\theta^{a} \equiv \frac{a^{a}}{f}$. \\
   
In the $SU(N-1)\times U(1)$-symmetric vacuum of infinite extent, the Goldstones are well defined everywhere. The mode-expansions for both Goldstones and $\xi$-particles are given by plane-waves. Of course, the finite size of the bubble affects the mode expansion and the profiles of various momentum eigenstates. In particular, the Goldstones exist only in the bubble interior and become ill-defined outside. The effective coupling that controls the scattering between the Goldstone modes of frequency $\omega$ and the $\xi$ quanta is given by $\alpha_G(\omega) = \frac{\omega}{f_{\omega}}$. The entire information about the mode profiles and other factors is encoded in the effective scale $f_{\omega}$. For the saturated bubble, $f_{\omega}$ is of the same order as the fundamental scale $f$. This is because for the saturated bubble all the scales are of the same order (\ref{OMR}). We thus have, $\alpha_G(\omega) = \frac{\omega}{f}$. This coupling controls all the Goldstone processes of our interest. \\
 
Since $\xi$-quanta are massless, the conservation of energy and of the $SU(N)$ charge are not an obstacle for the bubble decay. However, each elementary vertex involving $\xi$, is suppressed by the Goldstone coupling $\alpha_G$. Due to this, an explosive decay of the bubble into a large number of $\xi$ quanta is highly improbable. Such an ``explosion'' does not represent a significant channel of the bubble-decay. This is similar to the suppression of the explosive decay of a black hole into a large number of soft gravitons. \\   
   
The  leading order processes contributing to the bubble evaporation
are of the following two types. The first category includes the decays of Goldstones of frequency $\omega$ into pairs of the asymptotic $\xi$ quanta. The second category includes the re-scatterings of the pairs of Goldstones into $\xi$-s. The rates of these processes are given by, 
  \begin{equation}
  \Gamma_{1\rightarrow 2} \sim \omega \alpha_{G} N_G \sim
  1/R \,,
 \end{equation}  
 and 
 \begin{equation}
  \Gamma_{2\rightarrow 2} \sim \omega \alpha_{G}^2 N_G^2 \sim
  1/R \,,
 \end{equation}  
respectively. Here we took into account that the saturated bubbles satisfy the relation (\ref{OMR}), as well as, $N_G \sim N_{\text{Gold}} \sim 1/\alpha_{G}$. The re-scattering processes involving higher number of Goldstones are more suppressed. Thus, the decay rate of the saturated bubble is given by, 
   \begin{equation} \label{Rtotal}  
  \Gamma_{\text{decay}} \sim 
  1/R \,.
 \end{equation}  
This is nothing but the familiar Hawking rate of a black hole decay. Just like a black hole, on average, the saturon bubble emits a quantum of energy $\Delta E \sim 1/R$ per time $\Delta t \sim R$. Correspondingly, the power of the emitted radiation,
\begin{equation}  \label{Power}  
 P \,  \sim \, \frac{1}{R^2} \,\sim \,  T^2 \,, 
 \end{equation}  
is similar to the power of radiation from a black hole of temperature (\ref{Temperature}). \\

It is clear that the emission of $\xi$ quanta of energies $E \gg 1/R$, requires a re-scattering of a larger number $n$ of Goldstones into a small number of $\xi$-s. Such processes are highly suppressed. For example, production of $\xi$ quanta of energy $E = n/R$ with $n \gg 1$, requires re-scattering of $\sim n$ Goldstones. This brings an extra suppression factor, $\mathrm{e}^{-n}$, characteristic for the processes $n \rightarrow 2$ at large $n$ \cite{Dvali:2020wqi}. \\
  
Thus, the rate of emission of quanta of energy $E \gg 1/R$, is suppressed by an exponential factor of the type, 
 \begin{equation}\label{ERfactor}
   \Gamma_{E \gg T} \sim \mathrm{e}^{-ER} \,.
 \end{equation}   
  For an observer unfamiliar with the microscopic origin of the suppression, the factor (\ref{ERfactor}) is naturally interpreted and the Boltzmann factor 
  \begin{equation} \label{Bfactor}
   \Gamma_{E \gg T} \sim \mathrm{e}^{-\frac{E}{T}}\,,
 \end{equation} 
 characteristic of thermal radiation with the effective temperature $T \sim 1/R$, as given by (\ref{Temperature}). Such an observer would naturally conclude that the saturon state is thermal, despite the fact that the microscopic theory clearly tells us otherwise. The secret is in $1/N$ ($1/S$) corrections, which carry the information about the purity of the state. Only after clearly resolving these corrections, can an observer identify in which pure state the saturon is. However, such a resolution takes a rather long time, given by (\ref{Volume}). This aspect of radiation shall be quantified below.

\section{Information in Hawking radiation} 

Let us now discuss how the information is extracted from the bubble by the Hawking radiation. We start by noticing that in the semi-classical limit (\ref{Limit}) the quantum information carried by the radiated quanta is strictly unresolvable. This is obvious from the fact that in this limit all the quantum couplings vanish, $\alpha_G, \alpha_{\xi}, \alpha\,  \rightarrow \, 0$. It is therefore impossible for an external observer to recognize the $SU(N)$ quantum numbers of the outgoing radiation. At the same time, in the limit (\ref{Limit}) the evaporation rate remains finite and so does the power of the emitted radiation (\ref{Power}). This is at the expense of infinite $N_G$.
  
Thus, an observer sees a persistent radiation, with an effective temperature (\ref{Temperature}), but is unable to resolve its information content within any finite time. 
Hence, in the semi-classical limit (\ref{Limit}), the information processing by a saturon bubble is not any different from that of a black hole. \\
   
Let us discuss this crucial point more carefully. For this, we compare the evaporation of two saturated bubbles that carry distinct information patterns,  $\ket{n_{\omega}^1, n_{\omega}^2, \, \dotsc}$  and  $\ket{n_{\omega}^{\prime 1}, n_{\omega}^{\prime 2}, \, \dotsc}$, which are related by $SU(N)$-transformation. The patterns satisfy the constraint (\ref{ConstN}). \\
  
Notice, despite the fact that $SU(N)$ commutes with the Hamiltonian, the two bubbles carry the distinct quantum information messages. Since the information patterns are $SU(N)$-transformed relative to one another, they will interact differently with a fixed reference probe. \\
  
Let us recall that for a saturated bubble, the parameters satisfy the following important relations,
 \begin{equation}  \label{AAANNN}
  \alpha \sim \alpha_G \sim \alpha_{\xi}\, \sim 
  \frac{1}{N_{\text{Gold}}} \sim  \frac{1}{N_{G}}
   \sim  \frac{1}{N}\,.   
 \end{equation} 
 That is, for a saturated bubble, the total occupation number of Goldstone modes $N_G$ is of order the number of the Goldstone species $N_{\text{Gold}}$, and both numbers are of order $N$. At the same time, all of the collective couplings are critical. \\
  
If the bubble is saturated, the above relations hold regardless whether we work in the semi-classical approximation or in the full quantum theory. The difference is that in quantum theory, although the numbers 
are large (and all $\alpha$-couplings are weak), they are finite. In the exact semi-classical limit, all three couplings become infinitely weak, while numbers become infinitely large. \\
      
The two regimes (semi-classical versus quantum) of the saturated bubble can be summarized as, 
   \begin{equation}
  \alpha \sim \alpha_G \sim \alpha_{\xi}\, \sim 
  \frac{1}{N_{\text{Gold}}} \sim  \frac{1}{N_{G}}
   \sim  \frac{1}{N}\, = \, \begin{cases}
   \neq 0 \ll 1  & \text{in~full~quantum~theory}, \\
    =  0  & \text{in~semi-classical~theory}.
\end{cases}   
 \end{equation} \\
  
Now, let us compare the radiations coming from the above two bubbles. Since the two patterns are related by $SU(N)$-transformation, all the $SU(N)$-invariant characteristics of the decay process are identical for the two bubbles. For example, the two bubbles have the same decay rates, given by (\ref{Rtotal}). \\
    
In order to distinguish between the two bubbles, the observer must perform a measurement that can be sensitive to differences $\Delta n_{\omega}^{a}$ between the individual occupation numbers of the two patterns. In the semi-classical limit, this is only possible if some of the ratios $\Delta n_{\omega}^{a}/N_G$ remain finite. In other words, the observer can only differentiate among the patterns that are classically distinct. The relative fraction of such patterns vanishes in the limit of infinite $N_G$. The bulk of the quantum information is carried by the patterns in which the differences are small. Within a semi-classical treatment, such patterns cannot be told apart. This explains why in the semi-classical theory the radiation from the bubble carries no quantum information. Or to be more precise, the information is there but is unresolvable within any finite time. \\
   
Now, in full quantum treatment, the radiations coming out of the two bubbles with distinct information patterns, can be distinguished. Even if the differences $\Delta n_{\omega}^{a}$ are of order one, they are detectable. The distinction is encoded in the corrections of order $1/N_G$, or equivalently, of order $1/S$. \\ 
   
\begin{figure}[H]
\centering
\begin{tikzpicture}
\begin{feynman}
\vertex (L1) {\(\)};
\vertex[right=3cm of L1] (C1);
\vertex[right=3cm of C1] (H1);
\vertex[above=0.5em of H1] (R1) {\(\xi\)};

\vertex[below=1.5em of L1] (L2) {\(\)};
\vertex[below=1.5em of R1] (R2) {\(\xi^{\dagger}\)};
\vertex[below=1.5em of C1] (C2);

\vertex[below=1.5em of L2] (L3) {\(\)};
\vertex[below=2.8em of R2] (R3) {\(\)};
\vertex[below=1.5em of C2] (C3);

\vertex[below=1.5em of L3] (L4) {\(\)};
\vertex[below=1.5em of R3] (R4) {\(\)};
\vertex[below=1.5em of C3] (C4);

\vertex[below=0.8em of L4] (L5);
\vertex[below=0.8em of R4] (R5);
\vertex[below=0.15em of C4] (C5) {\(\vdots\)};

\vertex[below=1.5em of L5] (L6) {\(\)};
\vertex[below=1.5em of R5] (R6) {\(\)};
\vertex[below=1.5em of C5] (C6);

\vertex[below=1.5em of L6] (L7) {\(\)};
\vertex[below=1.5em of R6] (R7) {\(\)};
\vertex[below=1.5em of C6] (C7);
\diagram* {
{[edges=fermion]
(L1) -- [blue] (C1) -- [blue] (R1),
(R2) -- [red] (C2) -- [red] (L2),
(L3) -- (C3) -- (R3),
(R4) -- (C4) -- (L4),
(L6) -- (C6) -- (R6),
(R7) -- (C7) -- (L7),
},
};
\draw [decoration={brace}, decorate] (L7.south west) -- (L7.west |- L1.north)
node [pos=0.5, left] {\(\mathcal{B}~\)};
\draw [decoration={brace}, decorate] (R3.north east) -- (R3.east |- R7.south)
node [pos=0.5, right] {\(~\tilde{\mathcal{B}}\)};
\end{feynman}
\end{tikzpicture}
\caption{The decay $\mathcal{B} \to \tilde{\mathcal{B}} + \xi + \xi^{\dagger}$ as a planar diagram in 't Hooft notation.
The opposite arrows indicate the flow of $SU(N)$  ``color'' and anti-color. The  red and blue lines indicate the quantum numbers taken away by the emitted $\xi$ particles.}
\label{fig_Bdecay}
\end{figure}
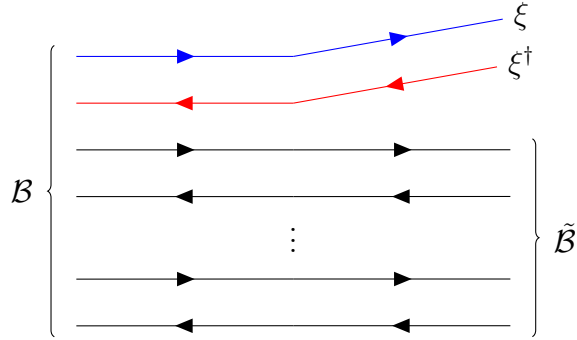

The process of the information emission is very transparent in the effective description of the bubble as of a 
composite state transforming as $SU(N)$ tensor representation (\ref{BWF}). The effective coupling between the bubble and the $\xi$ field has the following form, 
  \begin{equation}
   {\mathcal B}_{\alpha_1,\alpha_2, \, \dotsc \, ,\alpha_{N_T-1}, \, \alpha_{N_T}}^ 
   {\beta_1,\beta_2, \, \dotsc \, , \beta_{N_T-1}, \, \beta_{N_T}} \,
  \xi^{\dagger \alpha_{N_T}} \xi_{\beta_{N_T}}\, 
  \tilde{{\mathcal B}}^{\alpha_1,\alpha_2, \, \dotsc \, , \alpha_{N_T-1}}_{\beta_1,\beta_2, \, \dotsc \, , \beta_{N_T-1}} \,,
 \end{equation} 
where 
$\tilde{{\mathcal B}}^{\alpha_1,\alpha_2, \, \dotsc \, , \alpha_{N_T-1}}_{\beta_1,\beta_2, \, \dotsc \, , \beta_{N_T-1}}$ is the operator describing a bubble that transforms under a smaller representation of $SU(N)$. The emission process is given by 
\begin{equation}
   {\mathcal B}_{\alpha_1,\alpha_2, \, \dotsc \, ,\alpha_{N_T-1}, \, \alpha_{N_T}}^ 
   {\beta_1,\beta_2, \, \dotsc \, , \beta_{N_T-1}, \, \beta_{N_T}} 
 \rightarrow   
   \,
  \xi_{\alpha_{N_T}} \,  + \, \xi^{\dagger \beta_{N_T}}\, + \, 
  \tilde{{\mathcal B}}_{\alpha_1,\alpha_2, \, \dotsc \, , \alpha_{N_T-1}}^ 
   {\beta_1,\beta_2, \, \dotsc \, , \beta_{N_T-1}} \,,
 \end{equation} 
  We can describe the process as a planar diagram in 't Hooft's notation in which the oppositely directed arrows describe the flow of global $SU(N)$ ``color'' and ``anti-color'' (see figure \ref{fig_Bdecay}). This diagrammatic language makes the process of the outflow of information from the decaying bubble very clear. In particular, it is obvious that a single emission carries away a negligible fraction of the information stored in the bubble.  

\begin{figure}[H]
\centering
\begin{tikzpicture}
\begin{feynman}
\vertex (L1) {\(\)};
\vertex[right=3cm of L1] (C1);
\vertex[right=3cm of C1] (R1) {\(\xi\)};

\vertex[below=0.5em of L1] (L2) {\(\)};
\vertex[right=3cm of L2] (C2);
\vertex[right=3cm of C2] (R2);

\vertex[below=1.5em of L2] (L3) {\(\)};
\vertex[right=3cm of L3] (C3);
\vertex[right=3cm of C3] (R3);

\vertex[below=0.5em of L3] (L4) {\(\)};
\vertex[right=3cm of L4] (C4);
\vertex[right=3cm of C4] (R4);

\vertex[below=1.5em of L4] (L5) {\(\)};
\vertex[right=3cm of L5] (C5);
\vertex[right=3cm of C5] (R5);

\vertex[below=0.5em of L5] (L6) {\(\)};
\vertex[right=3cm of L6] (C6);
\vertex[right=3cm of C6] (R6);

\vertex[below=1.5em of L6] (L7) {\(\)};
\vertex[right=3cm of L7] (C7);
\vertex[right=3cm of C7] (R7);

\vertex[below=0.5em of L7] (L8) {\(\)};
\vertex[right=3cm of L8] (C8);
\vertex[right=3cm of C8] (R8);

\vertex[below=1.5em of L8] (L9) {\(\)};
\vertex[right=3cm of L9] (C9);
\vertex[right=3cm of C9] (R9);

\vertex[below=0.5em of L9] (L10) {\(\)};
\vertex[right=3cm of L10] (C10);
\vertex[right=3cm of C10] (R10) {\(\xi^{\dagger}\)};

\diagram* {
{[edges=fermion]
(L1) -- [blue] (C1) -- [blue] (R1),
(C2) -- (L2),
(L3) -- (C3),
(C4) -- (L4),
(L5) -- (C5),
(C6) -- (L6),
(L7) -- (C7),
(C8) -- (L8),
(L9) -- (C9),
(R10) -- [red] (C10) -- [red] (L10),
},
{[edges=horizontal]
(C3) -- [half right] (C2),
(C5) -- [half right] (C4),
(C7) -- [half right] (C6),
(C9) -- [half right] (C8),
},
};
\draw [decoration={brace}, decorate] (L10.south west) -- (L10.west |- L1.north)
node [pos=0.5, left] {\(\)};
\end{feynman}
\end{tikzpicture}
\caption{A diagram describing many $\rightarrow 2$ scattering, resulting in the emission of highly energetic $\xi$-s.}
\label{fig_Nto2decay}
\end{figure}
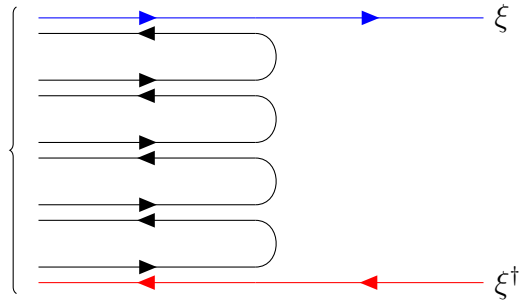

The diagrammatic language makes the origin of the exponential suppression factor (\ref{Bfactor}) very transparent. A typical process contributing in emission of highly energetic $\xi$-s is given by figure \ref{fig_Nto2decay}.

It is also clear that the back-reaction experienced by the bubble during the emission is of order $1/N$. Obviously, in the semi-classical limit (\ref{Limit}) the initial (${\mathcal B}$) and final ($\tilde{{\mathcal B}}$) bubble states are indistinguishable. At finite $N$, the difference 
is non-zero. However, the resolution takes time that scales with $N$. This determines the required time-scale for the information read-out, which we shall discuss next.

 \section{Time-scale of information retrieval} 
  
 The discussion of the information carried by Hawking radiation naturally brings us to the question of the time scale of information retrieval. In \cite{Dvali:2020wqi} it was argued that an arbitrary saturated system is subjected to an universal lower bound on the time-scale of information retrieval given by (\ref{Volume}). By applying the general argument of \cite{Dvali:2020wqi} to the present system, it is easy to see that the saturated bubble reproduces this expression. Indeed, let us consider such a bubble. As already discussed, the saturation is reached for the thick-wall bubbles, (\ref{OMR}), and they satisfy (\ref{AAANNN}). \\
        
As we already know, the quantum information is stored in the form of the memory pattern (\ref{pattern}) formed by the Goldstone modes. Therefore, if we wish to retrieve any quantum information from such a bubble, the interaction with the Goldstone modes is mandatory. The precise nature of the measurement is not important. The necessary condition for the extraction of the information is that the Goldstone modes must interact with some ``agents'' that can carry the $SU(N)$ quantum numbers outside the bubble. We can distinguish the two ways of information retrieval to which we shall refer to as \emph{passive} and \emph{proactive}. We shall discuss them separately. 
 
  \subsection{Passive retrieval} 
   
For the passive retrieval of information, an external observer has to analyze the properties of the Hawking radiation coming from the bubble. As we already discussed, to the zeroth order in $1/S$ (or $1/N$), the  radiation carries no information. In order to retrieve the information, the observer has to resolve $1/S$-corrections. For this, the observer must detect the $SU(N)$ content of the emitted quanta. This is a time-consuming process because of the following two reasons. \\
    
First, the quanta emitted at the initial stage carry a tiny fraction of the total information. For example, let us consider a decay of a particular Goldstone mode into a pair of $\xi$-s. In the language of the information pattern, the process can be described as a ``spontaneous emission'', during which one of the entries in the pattern (say $n_{\omega}^1$) decreases by one, 
   \begin{equation}
   \ket{n_{\omega}^1, n_{\omega}^2, \, \dotsc } \rightarrow 
 \ket{n_{\omega}^{1}-1, n_{\omega}^{2}, \, \dotsc} \, + \, 
 \xi^{\dagger} \, + \, \xi \,.
 \end{equation}  
   Of course, the $SU(N)$ charge of the decaying Goldstone is carried away by $\xi$-s. The typical wavelength of the emitted $\xi$ is $\sim R$.\\
    
Now, in order to  decipher the charge carried by $\xi$, the observer must take it through a detector with some sample particles. A maximally packed detector has the occupation number of probe quanta $N_{\xi} \sim 1/\alpha_{\xi}$ per de Broglie volume $\sim R^3$. The maximal interaction rate for an outgoing $\xi$-quantum with such a detector is, $\Gamma_{\xi} \sim \alpha_{\xi}^2 N_{\xi} /R \sim \alpha_{\xi}/R$. 
 Correspondingly, the minimal time required for recognizing the quantum numbers of $\xi$, is given by, 
 \begin{equation} \label{TX}
   t_{\xi} \sim  \dfrac{R}{\alpha_{\xi}} \sim SR \sim  R^3f^2\,.
  \end{equation} 
  However, the analysis of a single particle does not provide any significant knowledge about the information pattern carried by the bubble. In order to start gathering any reasonable amount of information about the $SU(N)$ charge content of the bubble, the observer needs to analyze at least of order $N_G$ emitted quanta. \\ 
   
However, the collection of such amount of quanta takes time, 
 \begin{equation}
   t_{\text{min}} \sim  N_G R \,, 
  \end{equation} 
 which is of the same order as (\ref{TX}). \\
 
 Thus, the picture is the following. In order to start getting any idea about the information content of the bubble, the observer needs to collect of order $N_G$ emitted quanta and also detect their $SU(N)$ quantum numbers. Both processes take the time (\ref{TX}) that is of the same order as 
 (\ref{Volume}).  
 
 \subsection{Proactive retrieval}  
 
 An observer has an alternative option for retrieving the quantum information stored inside the bubble. Instead of passively waiting for the bubble to evaporate, an observer can proactively scatter some probe quanta at the bubble and analyze the scattering products. As we shall see, also in this case the minimal time-scale for the start of the information retrieval is given by (\ref{Volume}). \\ 
   
The universal reason behind this is that $t_{\text{min}}$ cannot be shorter that the interaction time between the external probe and the Goldstone modes that carry the information pattern. Regardless of the setup, by Goldstone theorem, this interaction time is controlled by the Goldstone decay constant $f$. \\
   
The interaction rate is suppressed by the square of the Goldstone coupling $\alpha_G$, evaluated at the scale $\omega \sim 1/R$.  At the same time, the rate is enhanced by the total occupation number of Goldstones, $N_G$. Therefore, the rate is given by
\begin{equation}
  \Gamma_{\text{Gold}} \sim \alpha_{G}^2 m N_G \sim 
  \dfrac{1}{R^3f^2} \,.
 \end{equation}  
 The corresponding time-scale is given by,  
 \begin{equation}  \label{timeMB}  
t_{\text{min}} = \dfrac{1}{\Gamma_{\text{Gold}}}  \sim R^3f^2 \,. 
 \end{equation}  
This sets the minimal time of the start of information recovery by the proactive method. Due to the saturation relations, the time $t_{\text{min}}$ can be rewritten as, 
 \begin{equation}  \label{timeMB1}
t_{\text{min}} \sim SR \sim \dfrac{R}{\alpha} \,.  
 \end{equation}  
 We thus discover that with both methods (passive and proactive)  the minimal time for the start of the information retrieval is given by (\ref{Volume}). 
  
This result confirms the general point of \cite{Dvali:2020wqi} that saturated systems obey the universal lower bound (\ref{Volume}) on the minimal time of information-recovery. 
The black holes are by no means unique in this respect. For a non-gravitational saturon of size $R$, 
the time $t_{\text{min}}$ is equal to the Page time of a same-size black hole. 
\\
  
Notice, the expression (\ref{timeMB}) of information-recovery time nicely matches the existence of the information horizon in the semi-classical limit of the theory. From the expressions (\ref{Limit}) and (\ref{AAA3}), it is clear that in the same limit the information recovery time $t_{\text{min}}$ becomes infinitely long. Thus, the expressions (\ref{timeMB}) and (\ref{timeMB1}) explain why any saturated system 
must possess an information horizon in the semi-classical treatment of the theory. \\
   
Finally, we wish to comment that the above-discussed origin of Hawking radiation and of information-retrieval 
time-scales is very similar to the picture provided by the black hole $N$-portrait
\cite{Dvali:2011aa}. This is no accident, since the $N$-portrait describes a black hole as the saturated state of soft gravitons. By power of universality, the behaviour of the two systems  must be very similar. We shall comment more on the connection with black holes later.

\section{Numerical results}

For the numerical analysis we consider the case of occupying one $\theta^a$ macroscopically. For this, we take $\theta^a = \delta^{a1} \theta(x)$. Let us first write the Lagrangian (\ref{model_vac_bub}) in its full form in terms of the fields $\varphi$ and $\theta$,
\begin{equation}
\mathcal{L} = \dfrac{1}{2} \left(\partial_{\mu} \varphi \right) \left(\partial^{\mu} \varphi \right) + \dfrac{N}{4(N-1)} \varphi^2 \left(\partial_{\mu} \theta^a \right) \left(\partial^{\mu} \theta^a \right) - \dfrac{\tilde{\alpha}}{2} \varphi^2 \left(\varphi - \tilde{f} \right)^2 \, ,
\end{equation}
where 
\begin{equation} \label{tildAf} 
\tilde{\alpha} \equiv \alpha \frac{(N-2)^2}{N(N-1)}\,, ~~
\tilde{f} \equiv  f \frac{\sqrt{N(N-1)}}{(N-2)} \,.
 \end{equation}  
Now, let us define
\begin{equation}
\Psi \equiv \dfrac{1}{\sqrt{2}} \rho \text{e}^{i \chi / \tilde{f}}
\end{equation}
with
\begin{equation}
\rho = \varphi \, , \quad \chi = \sqrt{\dfrac{N}{2(N-1)}} \tilde{f} \theta.
\end{equation}
We can now re-write the Lagrangian as
\begin{equation}
\mathcal{L} = \left( \partial_{\mu} \Psi^* \right) \left( \partial^{\mu} \Psi \right) - \tilde{\alpha} \left| \Psi \right|^2 \left( \sqrt{2} \left| \Psi \right| - \tilde{f} \right)^2.
\end{equation}
The corresponding equations of motion for $\Psi$ are
\begin{equation} \label{eomPSI}
\Box \Psi + \tilde{\alpha} \Psi \left( \sqrt{2} \left| \Psi \right| - \tilde{f} \right) \left( 2 \sqrt{2} \left| \Psi \right| - \tilde{f} \right) = 0 \, .
\end{equation}
Solving these for $\Psi$, we can obtain the original fields $\varphi$ and $\theta$ as
\begin{equation}
\varphi=\sqrt{2}|\Psi| \quad \text{and} \quad \theta = \sqrt{\dfrac{2(N-1)}{N}} \text{Arg}(\Psi) \, ,
\end{equation}
respectively.

As initial conditions for the simulations we could consider
\begin{equation}
\Psi(t,r)|_{t=0} = \dfrac{\tilde{f}}{2\sqrt{2}} \left[ 1 + \tanh \left( \dfrac{ m(R_0-r)}{2} \right) \right]
\end{equation}
and
\begin{equation}
\partial_t \Psi(t,r)|_{t=0} = i  \tilde\omega \dfrac{\tilde{f}}{2\sqrt{2}} \left[ 1 + \tanh \left( \dfrac{ m(R_0-r)}{2} \right) \right],
\end{equation}
where $r$ is the radial coordinate, $R_0$ is the initial bubble radius, $\tilde\omega$ is the initial rotation frequency of the field $\chi$ and $m = \sqrt{\tilde{\alpha}}\tilde{f}= \sqrt{{\alpha}}{f}$. These can be equivalently expressed as
\begin{equation}
\varphi(t,r)|_{t=0} = \dfrac{\tilde{f}}{2} \left[ 1 + \tanh \left( \dfrac{ m(R_0-r)}{2} \right) \right]
\end{equation}
and
\begin{equation}
\dot{\theta} \equiv \partial_t \theta(t,r)|_{t=0} = \sqrt{\frac{2(N-1)}{N}} \tilde{\omega} \, ,
\end{equation}
respectively.

However, the above is valid only in the thin-wall approximation. Moreover, we are able to obtain better initial conditions for the simulation by first solving (\ref{radial}) numerically. We then set this solution as the initial condition for the simulations. This allows us firstly, to improve our numerical analysis, and secondly, to extend our analysis to that of thick-wall bubbles.

We numerically solve the equations of motion for $\Psi(t,r)$ in (3+1) space-time dimensions. In the simulations below we set $\tilde{\alpha}=1$ and $\tilde{f}=1$ throughout. From here on, just as we did before, we absorb the $N$-dependent pre-factors into the definitions of $\tilde{f}$, $\tilde{\alpha}$ and $\tilde{\omega}$ to obtain $f$, $\alpha$ and $\omega$, respectively. In our numerical simulations we find that for a certain \emph{critical frequency}, $\omega_c$, the bubble stabilizes, as expected, see figure \ref{fig_stable}. Thus, a certain critical macroscopic occupation number of the Goldstone modes, parametrized by $\omega_c$, stabilizes the vacuum bubble and impedes both its collapse and expansion. As expected, this is the same frequency as found in (\ref{thinwallR}). Below we shall re-derive this expression from the condition of conserved charge and, additionally, extend the calculation to obtain an estimate for the frequency of the bubble wall oscillations for a near-perfectly stabilized bubble.

\subsection{Critical frequency estimate}

In what follows we will assume spherical symmetry. Using the thin-wall approximation, the total energy of a vacuum bubble of radius $R \gg m^{-1}$ can be expressed as
\begin{equation} \label{TWEnergy}
E=\dfrac{2 \pi}{3 \alpha}m^3 R^2 (1-\dot{R}^2)^{-1/2} + \dfrac{2\pi}{3 \alpha} m^2 \omega^2 R^3 \, ,
\end{equation}
where we assumed $\omega=\omega(R)$ to be homogeneous in space.  In order to obtain an explicit expression for $\omega(R)$, we use the fact that the charge,
\begin{equation}
    Q=-i\int r^2 \, \mathrm{d}r \, \left(\Psi^*\partial_t\Psi - \Psi\partial_t\Psi^* \right) \, ,
\end{equation}
is conserved in time; $\dot{Q}=0$. In the thin wall approximation,
\begin{equation} \label{Charg}
Q= \dfrac{2\pi }{3} f^2 \omega R^3= \dfrac{2\pi }{3\alpha} m^2 \omega R^3 \, .
\end{equation}
Substituting this into (\ref{TWEnergy}) we can re-write the total energy of the bubble as
\begin{equation}
E=\dfrac{2 \pi}{3 \alpha}m^3 R^2 (1-\dot{R}^2)^{-1/2} + \dfrac{2\pi}{3\alpha} m^2 \left(\dfrac{3 \alpha Q}{2\pi m^2 R^3}\right)^2 R^3 \, .
\end{equation}
We are interested in the static bubble configuration.  Thus we take $\dot{R}=0$. The total energy therefore simplifies to
\begin{equation}
E=\dfrac{2 \pi}{3 \alpha}m^3 R^2 + \dfrac{3\alpha Q^2}{2\pi m^2 R^3} \, ,
\end{equation}
and solving $\frac{\mathrm{d} E}{\mathrm{d} R} =0$ for $R$, we find
\begin{equation}
    R_0 \equiv \left[ \dfrac{3}{2} \left( \dfrac{3\alpha Q}{2\pi} \right)^2 \right]^{1/5} m^{-1}\,.
\end{equation}
From here, using (\ref{Charg}), we obtain the following estimate for the critical frequency
\begin{equation}
\omega_c = \sqrt{\dfrac{2 m}{3 R_0}} \, .
\end{equation}
This is the same frequency as that in (\ref{thinwallR}). \\
  
We will set $R_0$ as the initial bubble radius as an initial condition in the numerical simulations below. The total energy and the total charge for this static bubble are
\begin{equation}
E_0 \equiv \dfrac{40 \pi  m^5}{81 \alpha \omega_c^4} \quad \text{and} \quad Q_0 \equiv \dfrac{16 \pi}{81 \alpha} \left( \dfrac{m}{\omega_c} \right)^5 \, ,
\end{equation}
respectively. In what follows we test the above estimate and investigate the response of the system in various frequency regimes. \\
  
We make a further note on the radius oscillations in the near-critical frequency regime. Namely, we can estimate their frequency, $\omega_0$, in the following way.  Let us take $R=R_0+\delta R$. Then the energy in (\ref{TWEnergy}) becomes
\begin{equation}
E=\dfrac{10 \pi  m^3 R_0^2}{9 \alpha}
+\left(\dfrac{\pi m^3 R_0^2}{3 \alpha }\right)\delta\dot{ R}^2 
+ \left(\dfrac{10 \pi m^3}{3 \alpha } \right)\delta R^2
+\mathcal{O}\left(\delta R^3, \delta R^2 \delta\dot{R}, \delta R \, \delta\dot{R}^2, \delta\dot{R}^3\right)\, .
\end{equation}
We can now extract the frequency of the radius oscillations from the above as
\begin{equation}
\omega_0 \equiv \sqrt{10} R_0^{-1} = \sqrt{10}\dfrac{3\omega_c^2}{2m} \, .
\end{equation}

\subsection{Critical frequency regime}

\begin{figure}[H]
     \centering
     \begin{subfigure}[b]{0.45\textwidth}
         \centering
         \includegraphics[width=\textwidth]{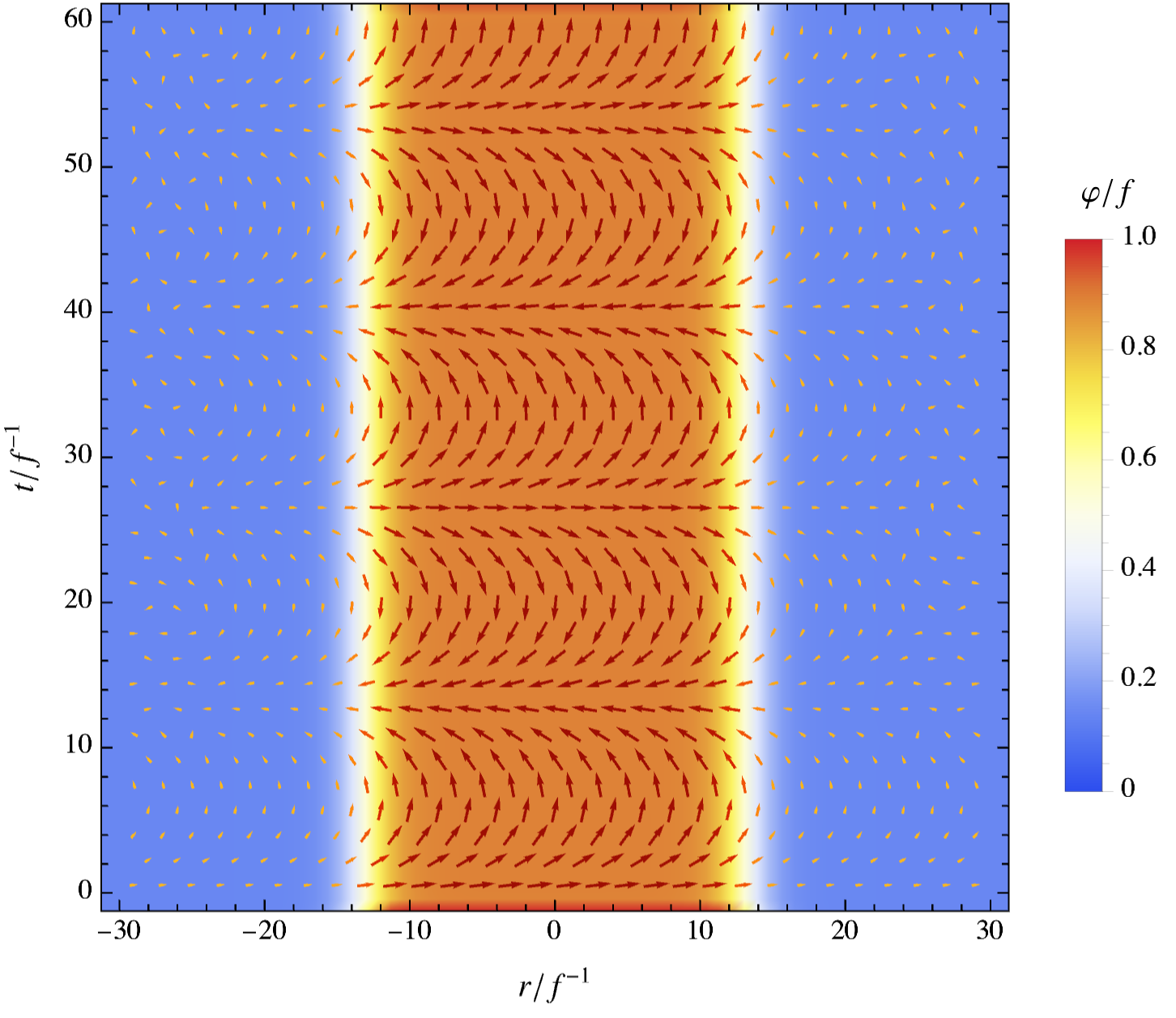}
         \caption{$R_0=12/f^{-1}$}
     \end{subfigure}
     \begin{subfigure}[b]{0.45\textwidth}
         \centering
         \includegraphics[width=\textwidth]{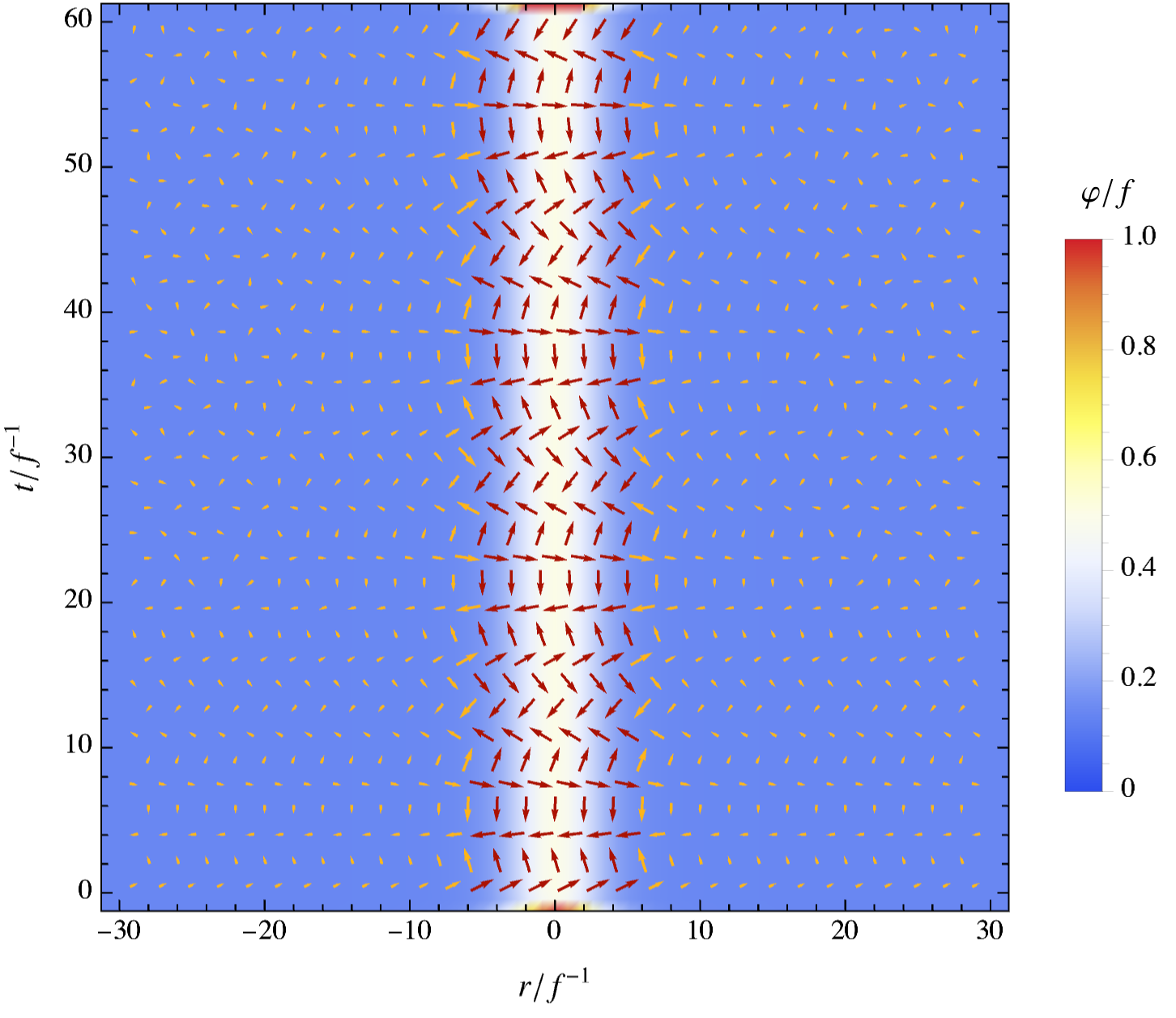}
         \caption{$R_0=1.01611/f^{-1}$}
     \end{subfigure}
        \caption{The frequency for both simulations is $\omega = \omega_c$.}
        \label{fig_stable}
\end{figure}

A certain critical macroscopic occupation number of the Goldstone modes, parametrized by $\omega_c$, stabilizes the vacuum bubble and impedes both its collapse and expansion (see figure \ref{fig_stable}). As we illustrate in the next section, bubbles in the low frequency regime, specifically with $\omega=0.1 \omega_c$, collapse at $t_{\text{col}} \sim R_0$. Notice that here, in the critical frequency regime, at $4 t_{\text{col}} \simeq 60 / f^{-1}$, the bubbles have not collapsed.

\begin{figure}[H]
     \centering
     \begin{subfigure}[b]{0.45\textwidth}
         \centering
         \includegraphics[width=\textwidth]{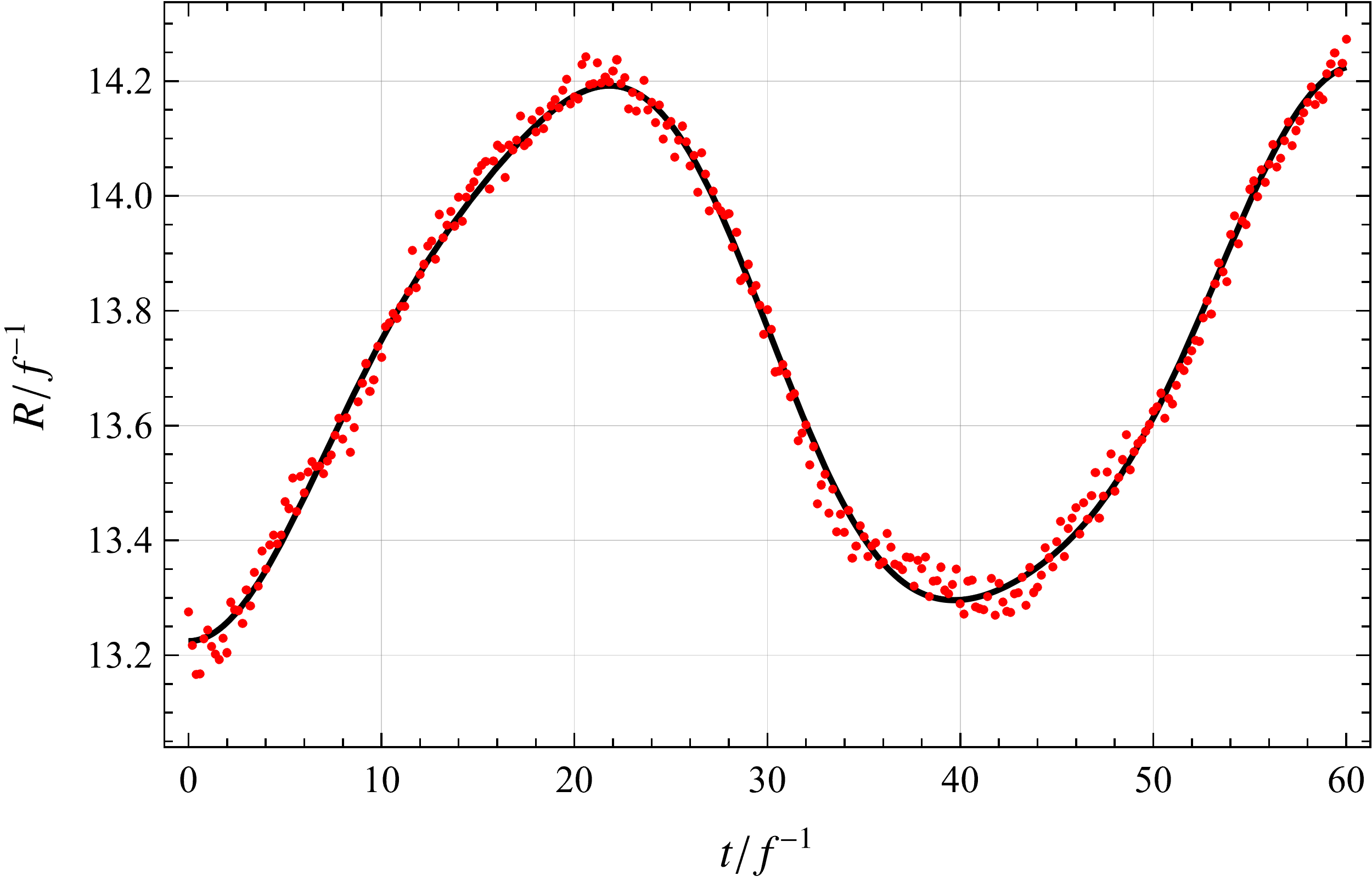}
         \caption{$R_0=12/f^{-1}$}
     \end{subfigure}
     \begin{subfigure}[b]{0.45\textwidth}
         \centering
         \includegraphics[width=\textwidth]{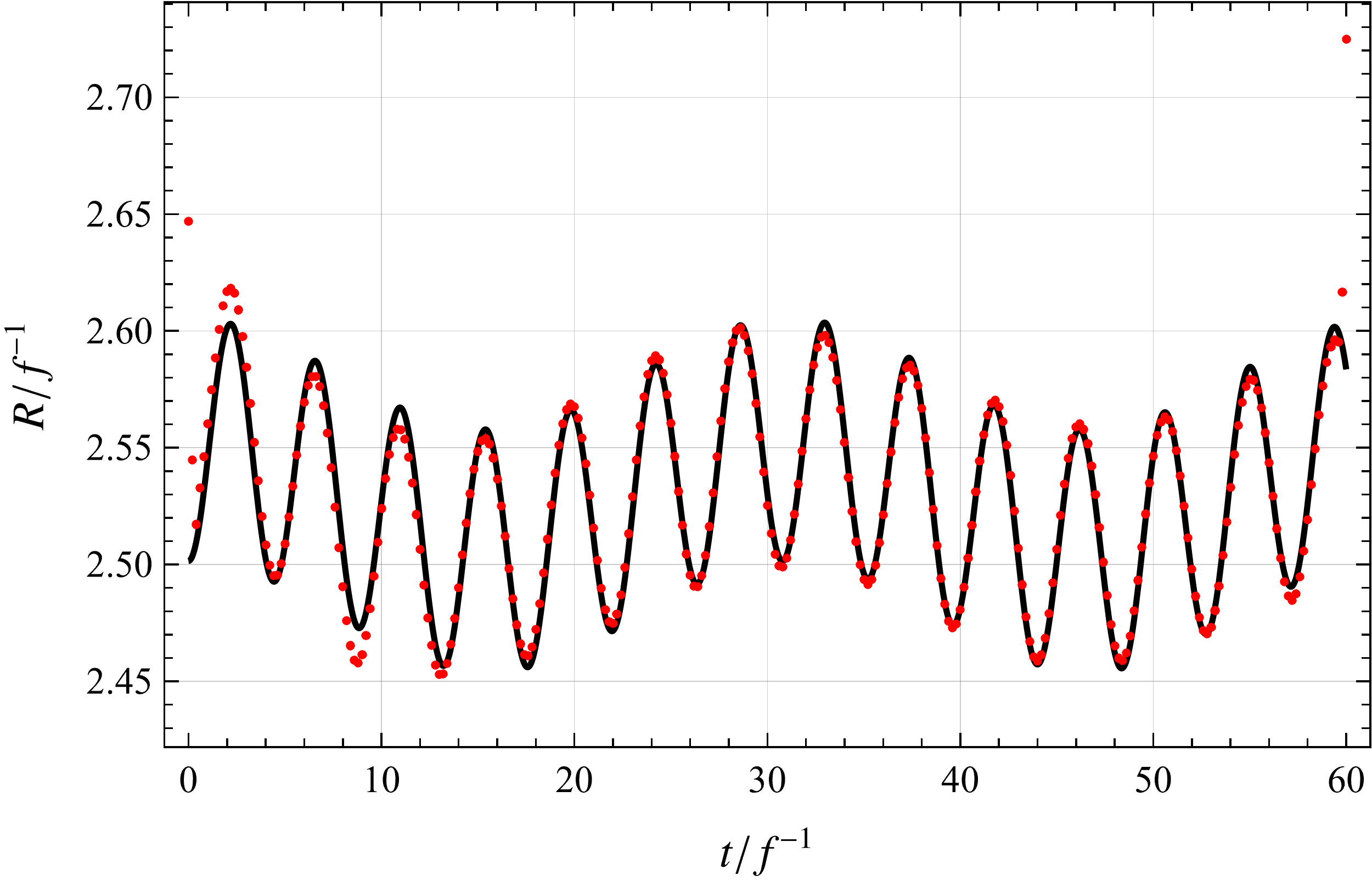}
         \caption{$R_0=1.01611/f^{-1}$}
     \end{subfigure}
     \caption{The frequency for both simulations is $\omega = 1.1 \omega_c$. The bubble radius, determined by the $r$-coordinate of the maximum of the energy density for each time step, is shown in red. The corresponding numerical fit $R(t)=R_m + A_0 \cos (\omega_0 t) + A_1 \cos (\omega_1 t)$ is shown in black.}
     \label{fig_osc}
\end{figure}

For a frequency slightly distinct from $\omega_c$, the bubble performs oscillations around its mean value $R_m$. We find at least two frequencies that contribute to the oscillations. Figure \ref{fig_osc} illustrates this.

Below we list the fit parameters of the fit function $R(t)=R_m + A_0 \cos (\omega_0 t) + A_1 \cos (\omega_1 t)$ in figure \ref{fig_osc}. For the large bubble these are: $R_m = 13.730 \pm 0.002$, $A_0 = -0.464 \pm 0.003$, $\omega_0 = 0.1537 \pm 0.0002$, $A_1 = -0.042 \pm 0.002$ and $\omega_1 = 0.370 \pm 0.002$. The coefficient of determination $\overline{R}^2$ and the unbiased root-mean-square error (RMSE), both adjusted for the number of fit-model parameters, are $0.999995$ and $0.0301$, respectively. For the small bubble these are: $R_m = 2.5297 \pm 0.0004$, $A_0 = -0.0519 \pm 0.0005$, $\omega_0 = 1.4287 \pm 0.0003$, $A_1 = -0.0237 \pm 0.0005$ and $\omega_1 = 0.2023 \pm 0.0006$, with $\overline{R}^2=0.999994$ and $\text{RMSE}=0.00599$. Four outlier data points were excluded from the fit for the small bubble. Both fit results of $\omega_0$ are close to our analytical estimate of $\omega_0=\sqrt{10}R_0^{-1}$ from before; $0.264$ and $3.11$ for the large and the small bubbles, respectively.

\subsection{Low frequency regime}

\begin{figure}[H]
     \centering
     \begin{subfigure}[b]{0.45\textwidth}
         \centering
         \includegraphics[width=\textwidth]{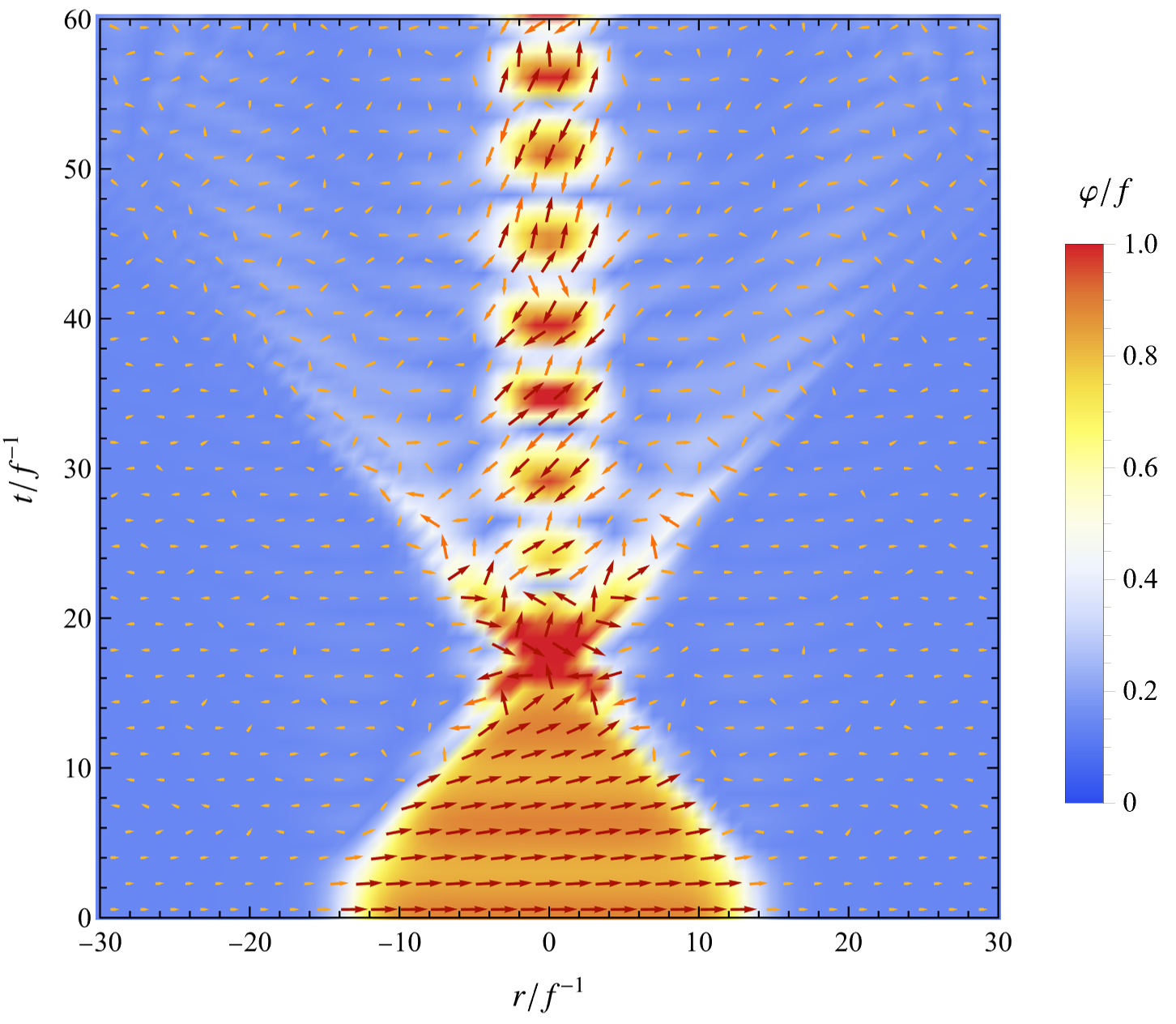}
         \caption{$R_0=12/f^{-1}$}
     \end{subfigure}
     \begin{subfigure}[b]{0.45\textwidth}
         \centering
         \includegraphics[width=\textwidth]{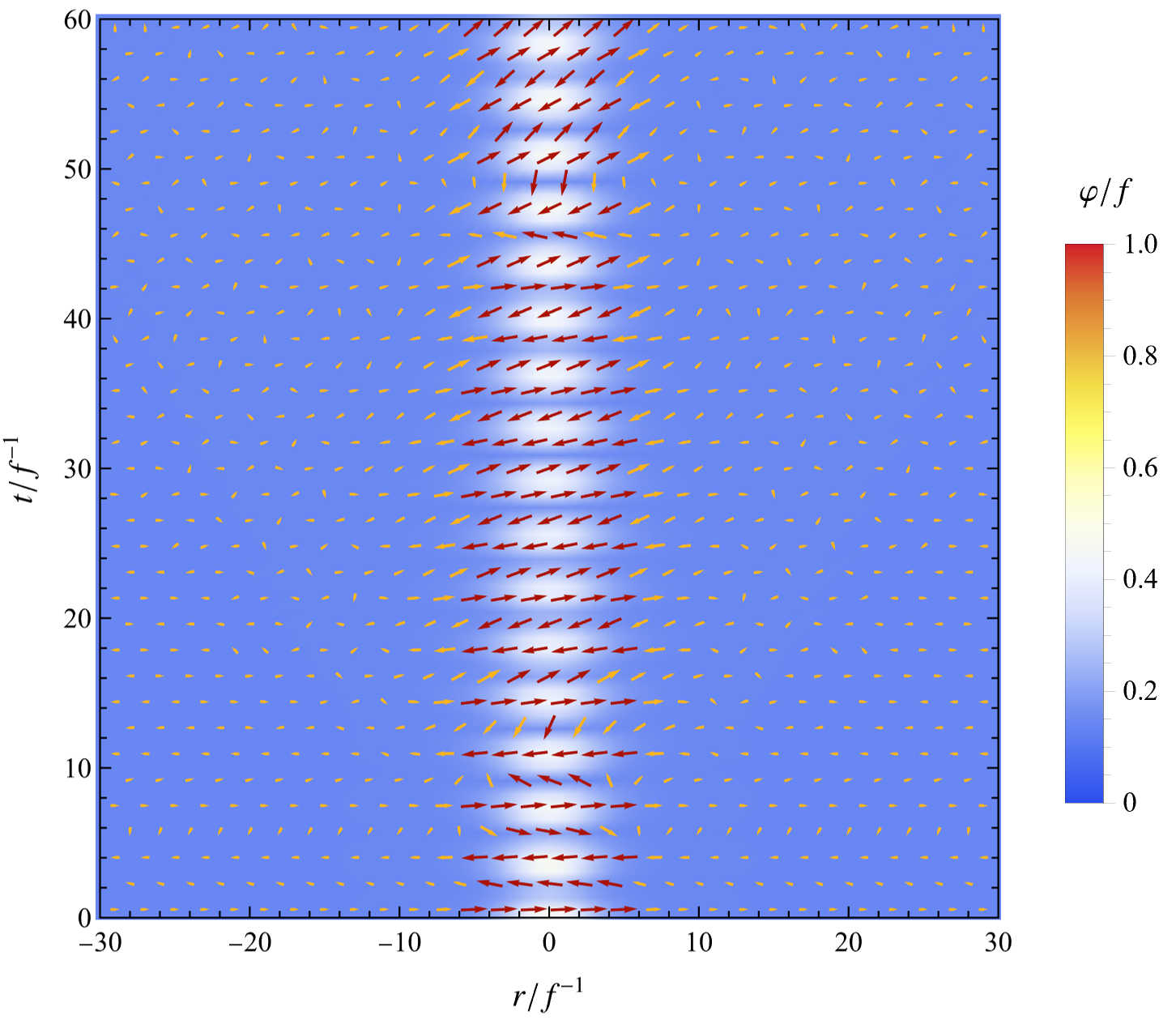}
         \caption{$R_0=1.01611/f^{-1}$}
     \end{subfigure}
        \caption{Time evolution of the bubbles with different initial radii $R_0$. The frequency for both simulations is $\omega = 0.1 \omega_c$. The color bar corresponds to $\sqrt{2}|\Psi(t,r)|/f=\varphi(t,r)/f$. The arrows represent the vector $\sqrt{2} \Psi(t,r)$ on the complex plane and thus indicate the value of $\theta(t,r)$.}
        \label{fig_collapse}
\end{figure}

For sufficiently low frequencies $\omega$ the bubble collapses. This is not surprising, as in this case the pressure due to the $\chi$ field's rotation is insufficient to counteract to the tension of the bubble wall. Thus, the vacuum bubble decays. For $\omega=0$ we numerically obtain that the bubble collapses. This regime has already been extensively studied in the literature. Here we consider the case $\omega=0.1 \omega_c$ and observe that the bubble collapses also in this case. From figure \ref{fig_collapse} we are able to read off the collapse time of the large bubble as $t_{\text{col}} \simeq  1.25 R_0 = 15/f^{-1}$. We estimate the collapse time of the small bubble also as $t_{\text{col}} \sim R_0$.

\subsection{High frequency regime}

\begin{figure}[H]
     \centering
     \begin{subfigure}[b]{0.45\textwidth}
         \centering
         \includegraphics[width=\textwidth]{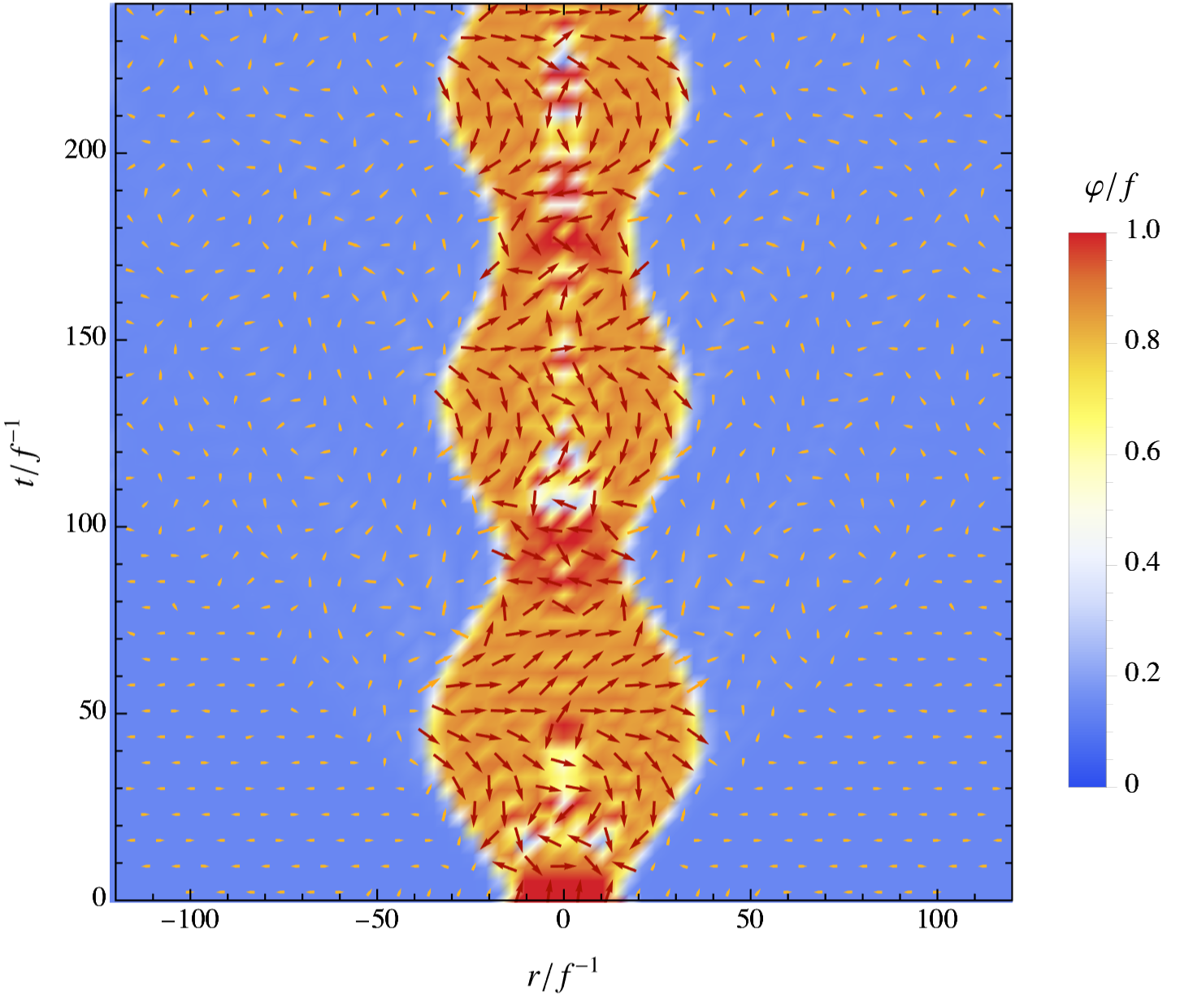}
         \caption{$R_0=12/f^{-1}$}
     \end{subfigure}
     \begin{subfigure}[b]{0.45\textwidth}
         \centering
         \includegraphics[width=\textwidth]{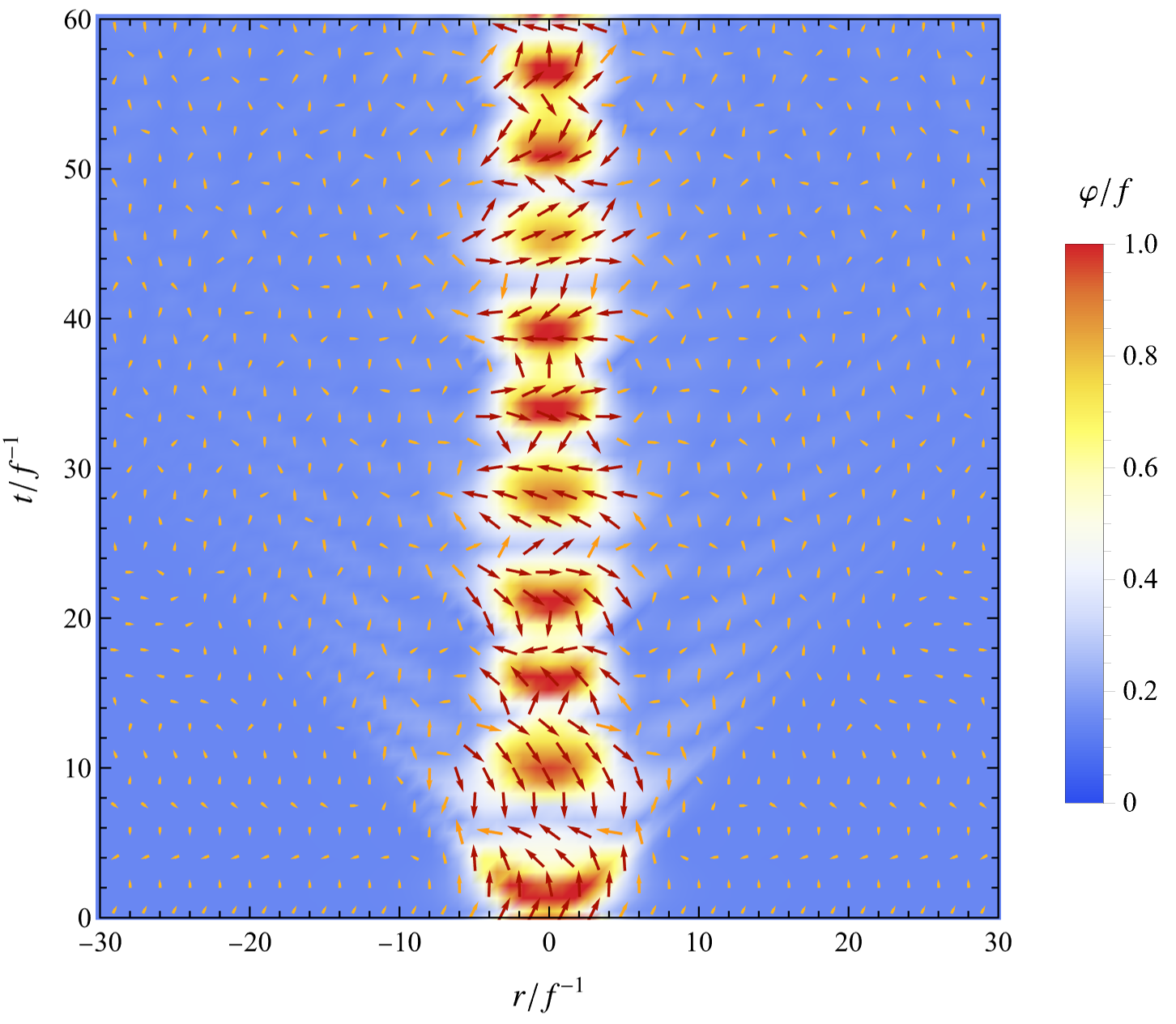}
         \caption{$R_0=1.01611/f^{-1}$}
     \end{subfigure}
        \caption{Time evolution of the bubbles with different initial radii $R_0$. The frequency for both simulations is $\omega = 4 \omega_c$. The color bar corresponds to $\sqrt{2}|\Psi(t,r)|/f=\varphi(t,r)/f$. The arrows represent the vector $\sqrt{2} \Psi(t,r)$ on the complex plane. Note the increased domain for the large bubble in panel (a).}
        \label{fig_exp}
\end{figure}

In the high frequency regime, the bubble initially expands: see figure \ref{fig_exp}. This is clear as the corresponding $\omega$ is above the critical value. Again, this should be intuitive, as in this case the pressure from the internal rotation is higher than the bubble wall tension. Once the bubble has expanded to a radius where the internal pressure is counterbalanced by the wall tension, it starts shrinking again. After this the process repeats itself. However, we additionally observe that a part of the energy is lost to the surroundings. This is by no means unexpected. In fact, this is a consistency check, which our numerical analysis fulfills. In the semi-classical limit the energy of the bubble is infinite. Therefore, any finite mass gap should result in waves propagating away from the oscillating bubble. This is exactly what we observe.

\subsection{Information horizon}

To illustrate the notion of the information horizon of the vacuum bubble, for the initial $r$-profile of $\theta$ we consider a perturbation,
\begin{equation}
p(r)=\exp \left[ \dfrac{i \pi}{2} \dfrac{f(r)}{f(0)} \right] \, ,
\end{equation}
with the probability density function $f(r)$ of a normal distribution,
\begin{equation}
f(r)=\dfrac{1}{\sigma \sqrt{2\pi}} \mathrm{e}^{-\frac{1}{2}\left( r/\sigma \right)^2} \, ,
\end{equation}
where we set the size of the perturbation as $\sigma \equiv 5/m \gg 1/m$.

\begin{figure}[H]
     \centering
     \begin{subfigure}[b]{0.45\textwidth}
         \centering
         \includegraphics[width=\textwidth]{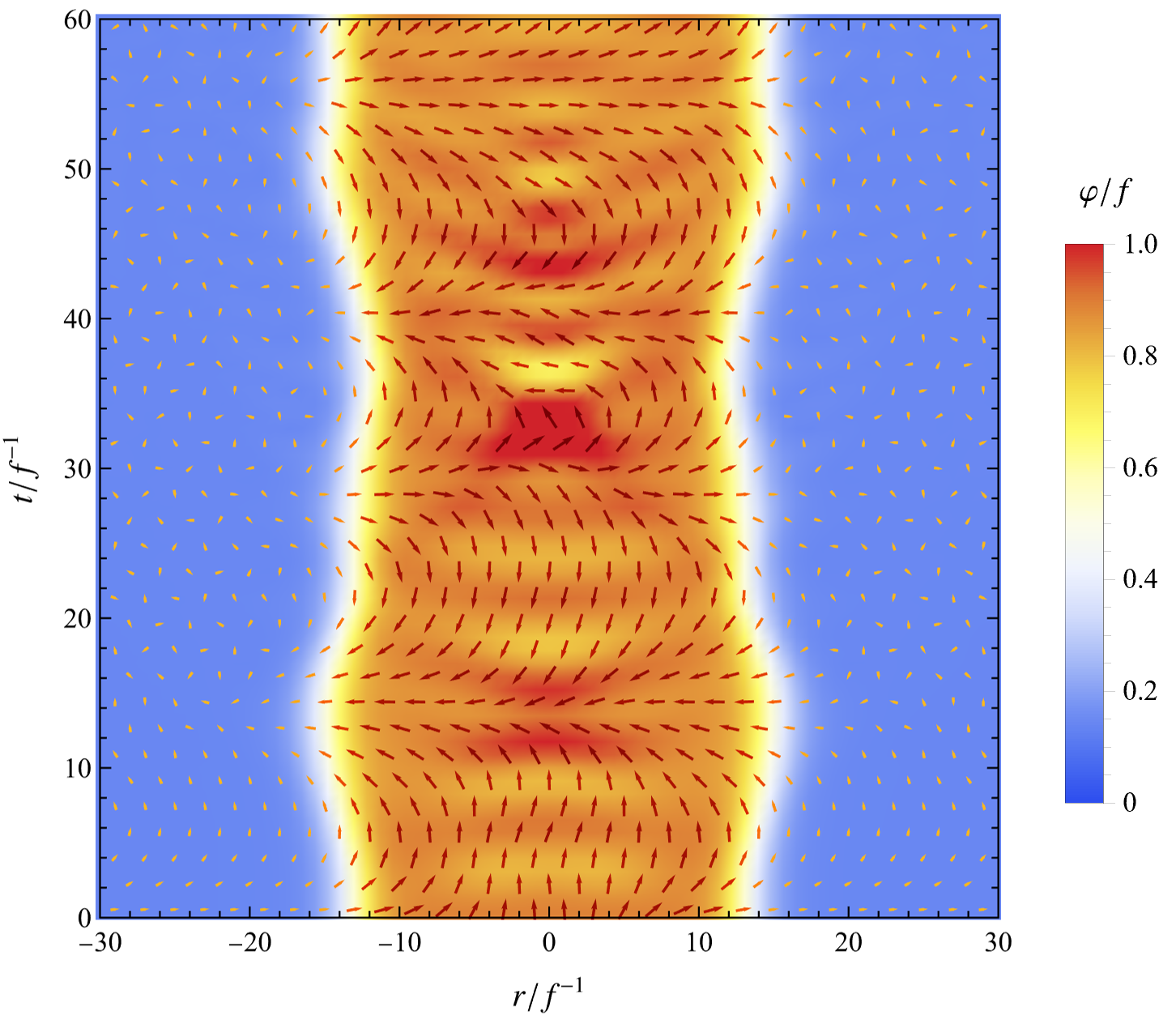}
         \caption{$R_0=12/f^{-1}$}
     \end{subfigure}
     \begin{subfigure}[b]{0.45\textwidth}
         \centering
         \includegraphics[width=\textwidth]{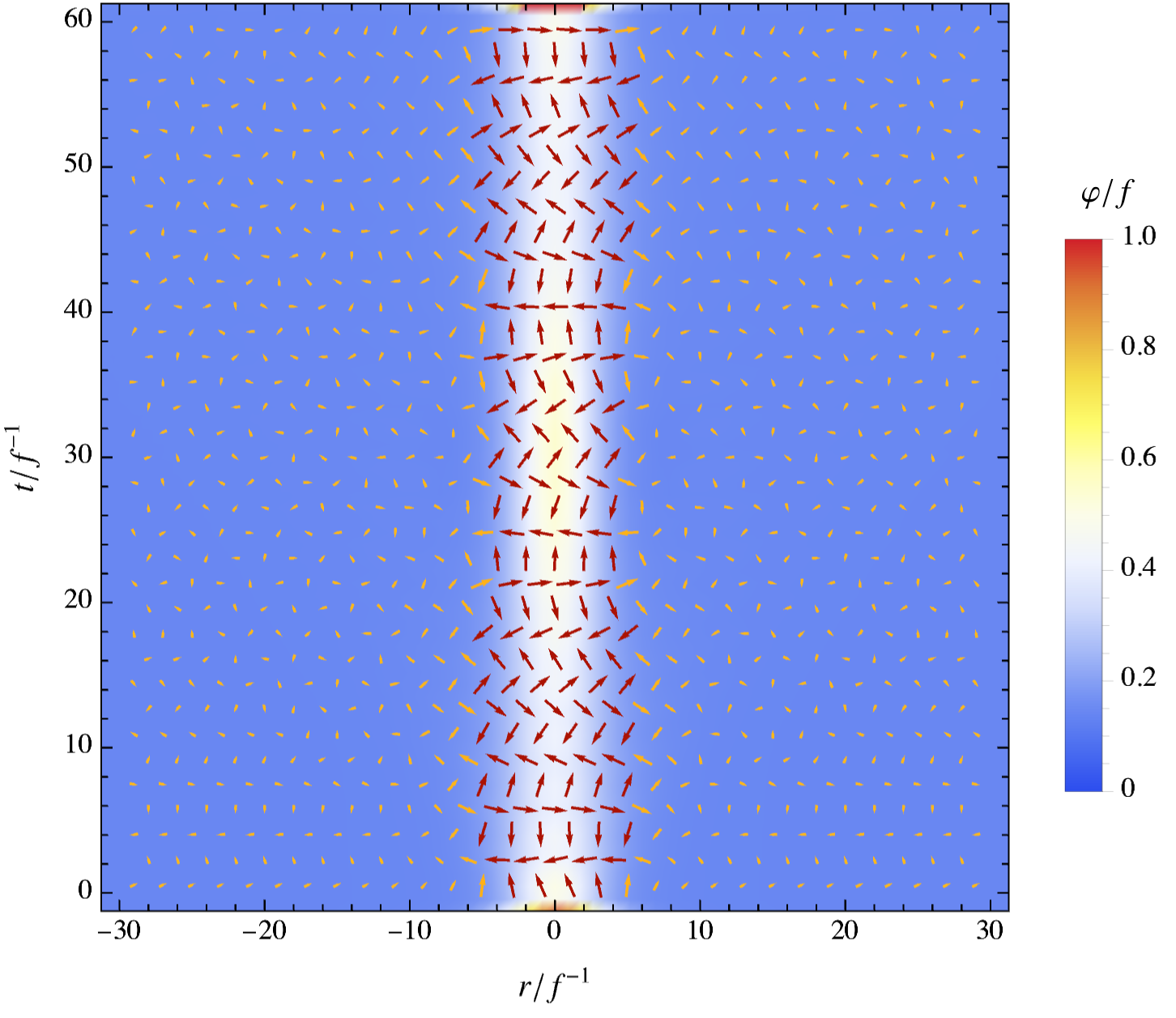}
         \caption{$R_0=1.01611/f^{-1}$}
     \end{subfigure}
        \caption{Time evolution of the bubbles with different initial radii $R_0$, with a perturbation of $\theta$. The frequency for both simulations is $\omega = \omega_c$. The color bar corresponds to $\sqrt{2}|\Psi(t,r)|/f=\varphi(t,r)/f$. The arrows represent the vector $\sqrt{2} \Psi(t,r)$ on the complex plane.}
        \label{fig_pert}
\end{figure}

Thus, the initial conditions are mapped from the old to the new ones in the following way: $\Psi|_{t=0} \to \Psi|_{t=0} p(r)$ and $\partial_t \Psi|_{t=0} \to \partial_t \Psi|_{t=0} p(r)$. Figure \ref{fig_pert} shows the corresponding simulation results. We observe that the total energy and the total charge are both conserved throughout the time evolution. Additionally, no waves are propagating out, in contrast to the high frequency regime. Moreover, the perturbation of $\theta$ is contained within the bubble. Additionally, note that $\theta$ is not homogeneous inside the bubble (cf.\ stabilized bubble in figure \ref{fig_stable}). The spatially dependent directions of the vector arrows clearly indicate the various values of $\theta$ within the bubble.

\section{Correspondence to black holes} 

We have seen that the saturated bubble exhibits a remarkable correspondence with a black hole, as suggested in \cite{Dvali:2020wqi}. The direct correspondence is 
established via the coupling $G_{\text{Gold}}^{(\text{P})}$
of the Goldstone boson of spontaneously broken 
Poincar\'{e} symmetry. When expressed through this parameter, the saturon of $SU(N)$ theory and a black hole in gravity share the same characteristics. Since the spontaneous breaking of Poincar\'{e} symmetry by any self-sustained object is an universal phenomenon, the coupling $G_{\text{Gold}}^{(\text{P})}$ is unambiguously defined in both cases. For a black hole, this quantity is equal to Newton's constant, whereas the decay constant $f$ is equal to Planck mass $M_{\text{P}}$. Once we understand this, it becomes clear that all the characteristics of a black hole 
are the same as of a generic saturon. This correspondence is summarized in table \ref{tab_sat_bh}. \\
     
\begin{table}[H]
\centering
\begin{tabular}{c|c|c}
\multicolumn{1}{c|}{\diagbox[innerwidth=3.2cm]{Quantity}{Object}} & Saturons & Black holes \\
\hline
$S $ & $(f R)^2=\alpha^{-1}$ & $(R M_{\text{P}})^2$ \\
$T$ & $R^{-1}$ & $R^{-1}$ \\
$t_{\text{min}}$ & $R^3 f^2=S R$ & $R^3 M_{\text{P}}^2 = S R$
\end{tabular}
\caption{Correspondence between properties of saturons and their specific realizations - black holes.}
\label{tab_sat_bh}
\end{table}
  
Using the above correspondence as the guiding principle, we can learn some useful lessons about black holes. First, we 
observe that $1/S$ ($1/N$) corrections to thermal radiation play the crucial role in purification of the state and in information recovery from it. Due to the same effects, the black hole evaporation is expected not to 
be self-similar, as opposed to the standard view. 
 The picture also makes it clear that during 
its decay the black hole state must be subjected to an inner entanglement, as it is suggested by the microscopic picture \cite{Dvali:2011aa}. 
 We also observe the phenomenon of the memory burden \cite{Dvali:2018xpy, Dvali_2020_2}
at work. We see that the saturated bubble is stable due to the quantum information that it carries.  \\
  
All of the above indicates that a black hole is likely a saturated bound state of gravitons, as this was suggested by the quantum $N$-portrait \cite{Dvali:2011aa}. The fact that we can formulate properties both for black holes and other saturons in terms of universal language of the Poincar\'{e} Goldstone, provides an additional support to this idea. The Goldstone boson of Poincar\'{e} symmetry shares one fundamental similarity with the graviton:  It is universally coupled to everything that carries energy-momentum. In case of a black hole, the two objects are the same. That is, the Poincar\'{e} Goldstone comes from the collective excitations of the graviton ``condensate''.   
   
 \section{Conclusions and outlook}

In the present paper we have verified 
the following points, proposed in \cite{Dvali:2020wqi}.
First, the unitarity upper 
bound on the entropy of a self-sustained object is given by (\ref{Area}), or equivalently,  by 
(\ref{alphaB}). 
 Secondly, the objects saturating this bound, called ``saturons'', 
 share the key properties with a black hole. The correspondence  is briefly 
 summarized in the table \ref{tab_sat_bh}.  
 The various aspects of this correspondence were already observed  
 on a number of examples, \cite{Dvali:2019jjw, Dvali:2019ulr, 
 Dvali:2021rlf}.  \\
 
 The present work represents a continuation of this program.   
   At the same time, we have incorporated 
  the study of another universal phenomenon
    exhibited by generic 
 systems of enhanced capacity of information storage. 
    This is a so-called  ``memory burden'' effect \cite{Dvali:2018xpy, Dvali_2020_2}. The essence of it is that the quantum information carried by an object tends to stabilize the ``host''.
  It has been suggested that due to its universal nature 
  the same effect must take place in black holes. 
  In the present paper we have detected this 
  phenomenon in a saturon of $SU(N)$ theory.  \\
   
 In order to address the above points, in the present work,  we have studied (both analytically and numerically) the 
 correspondence between black holes and the saturons
of a specific model, originally discussed in  
\cite{Dvali:2020wqi}. 
This is a renormalizable theory with 
$SU(N)$ global symmetry and coupling $\alpha$, 
which we took arbitrarily small.  
  The theory has a set  of degenerate vacua in which the $SU(N)$ symmetry is spontaneously broken down to various maximal subgroups.
As an asymptotic vacuum we chose the $SU(N)$-invariant one. 
In this theory we considered bubbles of
$SU(N-1)\times U(1)$ symmetric vacuum. 
 Due to spontaneous symmetry breaking, the bubble 
 interior houses $N_{\rm Gold} \sim N$ Goldstone species. \\

 We showed that the spectrum of the theory contains a tower of stationary bubbles.  
   In quantum description,  a bubble represents a bound state  
 of Goldstone modes with occupation number
 $N_G$. Bubbles that  minimize the energy cost of the Goldstone charge are stable in classical theory. 
 The stability can be understood 
 as a particular case of the ``memory burden'' effect. 
 The bubble is stable due to the quantum information encoded 
 in the $SU(N)$ flavor quantum numbers of the Goldstone modes that are ``locked up'' in the bubble interior. \\

   The stabilized bubbles can be viewed as generalized versions 
   of non-topological solitons, or $Q$-balls \cite{Lee:1991ax, Coleman:1985ki}. 
 However, the special property of the presented  bound state is 
 a very high entropy 
 carried by the Goldstone modes \footnote{~A different version 
 of a high entropy $Q$-ball was studied in \cite{GiaAndrei}.}. 
 This entropy reaches the limit (\ref{Area}), equivalently 
 (\ref{alphaB}), when the theory saturates the bound on unitarity.   A saturated bubble, acquires the properties of a black hole.
    These include: The area form of entropy; the information horizon; the thermal 
    decay with an effective temperature $T \sim 1/R$; and 
    the time-scale of information-retrieval similar to Page time.  
    All these characteristics, expressed in theory-invariant 
    quantities $G_{\rm Gold}$ or $\alpha_G$, have similar forms 
  for the saturated bubble and a black hole. \\
    
   Not surprisingly, close to saturation, the bubbles of $d=4$ dimensional  $SU(N)$ theory considered in the present work exhibit the close similarities with the recently found \cite{Dvali:2021rlf} saturon bound states in $d=2$ dimensional Gross-Neveu theory \cite{Gross:1974jv}.  
   This  close similarity is the manifestation of the universality 
of the phenomenon of saturation.  \\

   We have observed a certain deep connection between the
level of quantumness of the memory burden effect and the phenomenon of saturation.   
  For the stability of the bubble, only the total occupation 
    number of Goldstone modes $N_G$ matters.
  The memory burden effect depends only on 
 $N_G$ 
   and is insensitive to the specific occupation number distribution among the various 
  Goldstone species.  
    For example, a single Goldstone mode can be occupied macroscopically, as opposed to occupying multiple Goldstone modes microscopically.   \\   
       
 When the  number $N_G$ exceeds the number of Goldstone 
 species $N_{\rm Gold}$, the memory burden effect has a valid classical description. 
    This is because for $N_G \gg N_{\rm Gold}$,  necessarily, 
    some of the modes are occupied macroscopically.      
    We have observed that such bubbles always have less entropy than given by (\ref{Area}) and (\ref{alphaB}). 
    That is, the bubbles stabilized by a classical memory burden 
    are undersaturated. \\
    
     Conversely, the bubbles with $N_G \sim N_{\rm Gold}$
 can attain the maximal entropy (\ref{Area}).   
 For such bubbles we get a curious situation.
 They form an $SU(N)$ multiplet in which 
 the bubbles with classical memory burden are fully degenerate 
 with the quantum ones.  \\
 
   The construction of the present paper can be straightforwardly generalized to models with other field content and symmetry groups. For instance, $SO(N)$ with 
   symmetric representation has almost identical structure. 
   Also, it can be easily supersymmetrized.  \\

    In all known examples, the saturons represent 
  the bound states in which the occupation number of 
  quanta, their inverse couplings and entropy are of the same order.  Thus, the correspondence between the black holes and 
 saturons naturally hints towards the composite picture of 
 a black hole \cite{Dvali:2011aa}. 
 The bound state of $N$ gravitons describing a black hole in this
 picture is very similar to the saturated bound state of
$N$ Goldstones of $SU(N)$ theory.  
Many properties predicted for black holes by
$N$-portrait \cite{Dvali:2011aa}
are explicitly seen in the presented theory of saturons.  
 In particular, we saw how the information is carried away by $1/S$ effects and how from these effects the  
 time-scale (\ref{timeMB1}) of information-retrieval 
 emerges. \\

The discussed correspondence also reinforces 
 some previous suggestions for black holes. In particular, 
 the reality of the memory burden effect.  It is also evident that the 
 standard assumption about self-similar evaporation 
 of a black hole deserves to be reconsidered.




%
%
%
\end{document}